\documentclass{raa_twocolumn}            

\usepackage{graphicx,times,amsmath,amssymb}             
\usepackage{natbib}

\usepackage{pdflscape}
\usepackage{longtable}

\usepackage{booktabs}
\usepackage{multirow}
\usepackage{hyperref}
\hypersetup{colorlinks=true, linkcolor=blue, anchorcolor=red, citecolor=blue, filecolor=red, urlcolor=blue}

\begin{document}

\defcitealias{Hillenbrand2022AJ....163..263H}{H22}

\title{Variability of Young Stellar Objects in the Perseus Molecular Cloud}

\volnopage{Vol.0 (20xx) No.0, 000--000}      
\setcounter{page}{1}          

\author{Xiao-Long Wang\inst{1,2}, Min Fang\inst{1,3}, Gregory J. Herczeg\inst{4,5}, Yu Gao\inst{1,6}, Hai-Jun Tian\inst{7}, Xing-Yu Zhou\inst{4,5}, Hong-Xin Zhang\inst{3,8}, Xue-Peng Chen\inst{1,3}}

\institute{
Purple Mountain Observatory, Chinese Academy of Sciences, No. 10 Yuanhua Road, Qixia District, Nanjing 210023, People's Republic of China; {\it xlwang@pmo.ac.cn, mfang@pmo.ac.cn}\\
\and
University of Chinese Academy of Sciences, No. 19(A) Yuquan Road, Shijingshan District, Beijing 100049, People's Republic of China\\
\and
School of Astronomy and Space Science, University of Science and Technology of China, Hefei, Anhui 230026, People's Republic of China\\
\and
Kavli Institute for Astronomy and Astrophysics, Peking University, Yiheyuan Road 5, Haidian District, Beijing 100871, People's Republic of China\\
\and
Department of Astronomy, Peking University, Yiheyuan Road 5, Haidian District, Beijing 100871, People's Republic of China\\
\and
Department of Astronomy, Xiamen University, Xiamen, Fujian 361005, People's Republic of China\\
\and
School of Science, Hangzhou Dianzi University, Hangzhou, 310018, People's Republic of China\\
\and
Key Laboratory for Research in Galaxies and Cosmology, Department of Astronomy, University of Science and Technology of China, Hefei, Anhui 230026, People's Republic of China
}

\date{Received~~2009 month day; accepted~~2009~~month day}

\abstract{We present an analysis of 288 young stellar objects (YSOs) in the Perseus Molecular Cloud that have well defined $g$ and $r$-band lightcurves from the Zwicky Transient Facility. Of the 288 YSOs, 238 sources (83\% of our working sample) are identified as variables based on the normalized peak-to-peak variability metric, with variability fraction of 92\% for stars with disks and 77\% for the diskless populations. These variables are classified into different categories using the quasiperiodicity ($Q$) and flux asymmetry ($M$) metrics. Fifty-three variables are classified as strictly periodic objects that are well phased and can be attributed to spot modulated stellar rotation. We also identify 22 bursters and 25 dippers, which can be attributed to accretion burst and variable extinction, respectively. YSOs with disks tend to have asymmetric and non-repeatable lightcurves, while the YSOs without disks tend to have (quasi)periodic lightcurves. The periodic variables have the steepest change in $g$ versus $g-r$, while bursters have much flatter changes than dippers in $g$ versus $g-r$. Periodic and quasiperiodic variables display the lowest variability amplitude. Simple models suggest that the variability amplitudes of periodic variables correspond to changes of the spot coverage of 30\% to 40\%, burster variables are attributed to accretion luminosity changes in the range of $L_{\rm acc}/L_{\star}=0.1-0.3$, and dippers are due to variable extinction with $A_{V}$ changes in the range of $0.5-1.3$\;mag.
}

\authorrunning{X.-L. Wang, et al.}            
\titlerunning{YSO Variability, ZTF, LAMOST}                    

\maketitle

%
%
\section{Introduction}           

Photometric variability was one of the original defining characteristic of young stellar objects (YSOs), even before the sources were known to be young \citep{Joy1945ApJ...102..168J,Joy1946PASP...58..244J}. Different components of the YSO system (star + disk) dominate different part of the spectral energy distribution (SED) of the YSO, so monitoring at different wavelength probes the physical processes in different parts of the system \citep{Venuti2021AJ....162..101V,Fischer2022arXiv220311257F}. Optical monitoring is powerful for understanding the stellar rotation of spots of the stellar photosphere, accretion from the disk onto the star, and dust obscuration \citep[e.g.,][]{Hillenbrand2022AJ....163..263H,Cody2022AJ....163..212C}. Monitoring in the near- and mid-infrared bands has been used to study the warm dust in the disk, including the inner rim \citep[e.g.,][]{Skrutskie1996AJ....112.2168S,Carpenter2001AJ....121.3160C,Makidon2004AJ....127.2228M,MoralesCalderon2011ApJ...733...50M,Rebull2014AJ....148...92R,Park2021ApJ...920..132P}. These studies reveal higher fraction of variables for YSOs than for main-sequence stars, and that disked YSOs are more variable than diskless YSOs.

Time series photometry have revealed a diversity of lightcurve shapes, including dipping stars exhibiting episodic or quasiperiodic fading events \citep[e.g.,][]{Alencar2010A&A...519A..88A,Bodman2017MNRAS.470..202B}, bursting stars exhibiting discrete brightening events \citep[e.g.,][]{Stauffer2014AJ....147...83S}, and periodic variables displaying sinusoidal-like lightcurves. Many lightcurves have complicated shapes, with more than one potential phenomenon shaping the changes on many timescales. \citet{Cody2014AJ....147...82C} defined the flux asymmetry ($M$) and quasiperiodicity ($Q$) metrics to classify regularly sampled lightcurves from space-based observations into 7 categories: periodic, dipping, bursts, quasi-periodic, stochastic, and long-timescale. There are also other schemes classifying YSOs into different variability categories \citep[see Section~5.1 of ][for a review]{Cody2014AJ....147...82C}. In this work, we use the classification scheme of \citet{Cody2014AJ....147...82C} to separate the lightcurves into different categories.

Various mechanisms are responsible for the diversity of lightcurve shapes. Strictly periodic objects are attributed to rotational modulation due to the presence of star spots on the stellar surface, rotating into and out of view. The variability of dipping stars (both quasiperiodic and aperiodic dippers) is commonly explained as stemming from variable extinction due to time-dependent occultation by circumstellar material \citep{Alencar2010A&A...519A..88A,MoralesCalderon2011ApJ...733...50M,Ansdell2016ApJ...816...69A,Turner2014prpl.conf..411T}. Burst variables tend to display strong H$\alpha$ emission and red infrared colors \citep{Cody2018AJ....156...71C}, and their variability are related to accretion bursts \citep{Stauffer2014AJ....147...83S}. Stochastic lightcurves are likely to arise from continuously stochastic accretion behavior producing transient hot spots \citep{Stauffer2016AJ....151...60S}. Quasiperiodic behavior is generally interpreted as purely spot behavior on top of longer timescale aperiodic changes or a single variability behavior varies from cycle to cycle \citep{Cody2014AJ....147...82C}. The most probable mechanisms driving long timescale variability include variable extinction and variable accretion activity \citep{Parks2014ApJS..211....3P}.

Time series photometry from the Zwicky Transient Facility \citep[ZTF;][]{Kulkarni2018ATel11266....1K} has been used to study large samples of periodic variables \citep{Chen2020ApJS..249...18C}, as well as to investigate the variability behavior in YSOs \citep[][hereafter H22]{Hillenbrand2022AJ....163..263H}. In this paper, we analyze the variability properties of YSOs in the Perseus molecular cloud, using the time series photometry from the ZTF. The dataset and the sample are described in Section~\ref{sec:data}. The properties of the targets are determined in Section~\ref{sec:star_disk_acc}. In Section~\ref{sec:lc_analysis} we present the analyses of the lightcurves, the variability properties of our sample, and the CMD pattern. A discussion relating CMD patterns to simple models is presented in Section~\ref{sec:CMDpattern_model}. We give our summary in Section~\ref{sec:summary}.

\section{Data Set and Target Selection}
\label{sec:data}

\subsection{YSO Catalog}
\label{sec:data:YSO}

\begin{figure*}[!t]
    \centering
    \includegraphics[width=\textwidth]{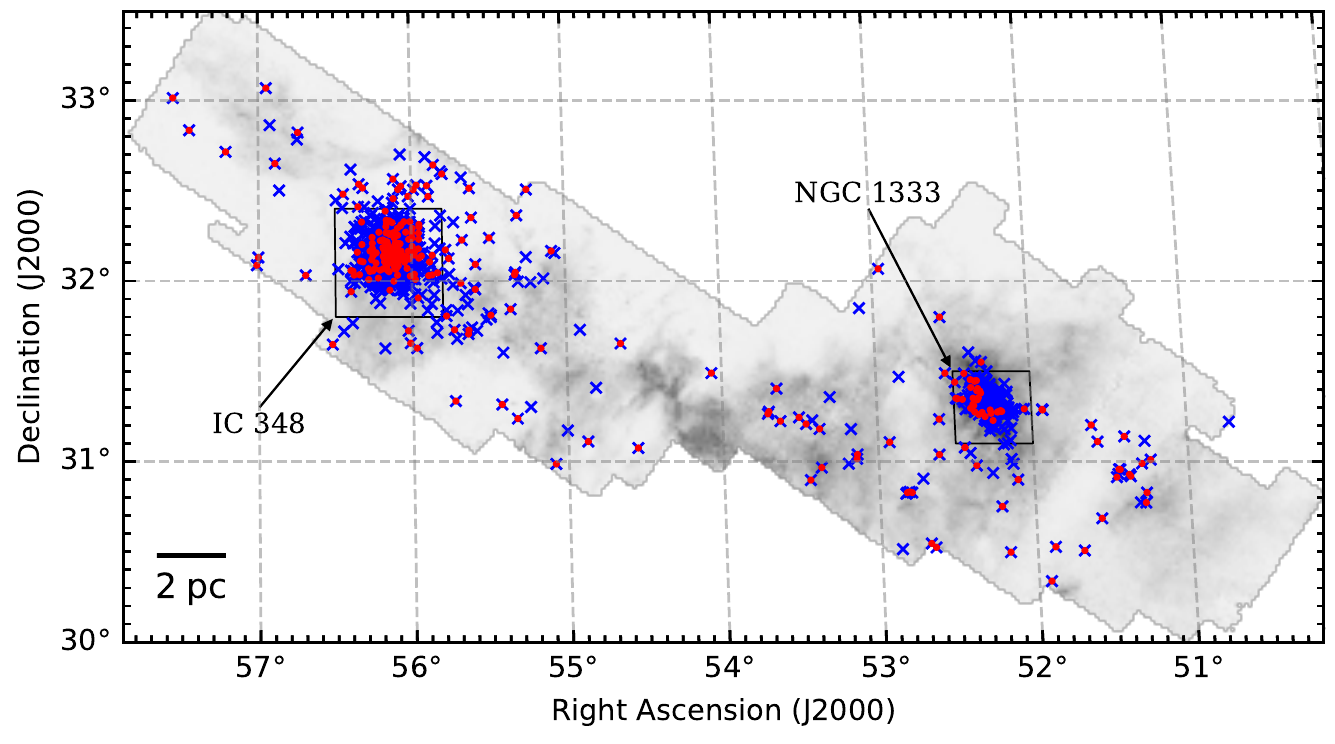}
    \caption{Spatial distribution of the initial YSO sample (blue crosses) overlaid on the FCRAO $^{12}$CO $J=1\to0$ integrated intensity map \citep{Ridge2006AJ....131.2921R}. Additional red points mark the 311 sources in our working sample. The two rectangles mark the two young clusters IC~348 and NGC~1333. The scale bar on the lower left shows a size of 2\;pc at a distance of 300\;pc.}
    \label{fig:YSOdistribution}
\end{figure*}

In our previous work \citep{Wang2022ApJ...936...23W}, we collected a sample of 805 previously known members from various literature \citep[i.e.,][]{Luhman2016ApJ...827...52L,Esplin2017AJ....154..134E,Luhman2020AJ....160...57L,Kounkel2019AJ....157..196K} and identified 51 new members based on Gaia astrometry \citep{Gaia-Collaboration2021A&A...649A...1G,Riello2021A&A...649A...3R,Fabricius2021A&A...649A...5F} and LAMOST spectroscopy \citep{Luo2022yCat.5156....0L}, resulting in a total of 856 well confirmed members in the Perseus molecular cloud. This sample of 856 members constitutes our initial YSO sample. The spatial distribution of the initial sample as well as our working sample (discussed below) are displayed in Figure~\ref{fig:YSOdistribution}.

\subsection{ZTF Photometry}
\label{sec:data:ZTF}

The Zwicky Transient Facility \citep[ZTF;][]{Kulkarni2018ATel11266....1K} is a time-domain photometric survey currently in progress. It uses a 47\;deg$^{2}$ camera consisting of 16 individual CCDs each 6k$\times$6k covering the full focal plane of the Palomar 48-inch (P48) Schmidt Telescope at Palomar Observatory \citep{Masci2019PASP..131a8003M}. In this paper, we analyze data from the thirteenth public ZTF data release (ZTF DR13\footnote{\url{https://irsa.ipac.caltech.edu/data/ZTF/docs/releases/dr13/ztf_release_notes_dr13.pdf}}), which corresponds to more than four years of data taken between 2018 March 17 and 2022 July 8 (58194$\leq$MJD$\leq$59768). The photometry is provided in the $g$, $r$ and $i$ bands, with a uniform exposure time of 30\;s in the public survey and is calibrated to the PanSTARRS photometry and reported in AB magnitude. The ZTF DR13 contains about 4.4 billion lightcurves in the $g$, $r$ or $i$ bands, with more than half of them have $\ge$20 epochs of observations. The $r$-band have the most number of lightcurves.

 Searching the ZTF archive, we extract $g$-band lightcurves for 466 members and $r$-band lightcurves for 577 members of the Perseus molecular cloud. In this work, we will focus our analysis on the ZTF $g$ and $r$-band lightcurves only, since no $i$-band lightcurve is available for our targets. We ignore observations with \texttt{catflags=32768} that are affected by clouds or contaminated by the moon. For our analysis, we restrict ourselves to sources with mean magnitudes brighter than 20.8 and 20.6 in the $g$ and $r$-band respectively\footnote{These magnitudes correspond to the median 5$\sigma$ sensitivity in 30\;s of $g$ and $r$-bands, respectively \citep{Bellm2019PASP..131a8002B,Masci2019PASP..131a8003M}.}, over the entire time series. Following the procedure in \citepalias{Hillenbrand2022AJ....163..263H}, we further remove observations taken on MJD days 58786, 58787, 58788, 58789 and 58805, that are part of the ZTF high-cadence experiments \citep{Kupfer2021MNRAS.505.1254K} and affect the period search. To alleviate the impact from potential outlier measurements in the lightcurves, we remove measurements 5$\sigma$ away from the median magnitude of the corresponding lightcurve. In some cases, an additional one to three points are found to be nonphysically discrepant and are removed. Finally, only lightcurves (5$\sigma$ clipped) with more than 30 measurements are considered for further analysis. Several sources which are located closely are not resolved by the ZTF are removed. In the current work, we only focus on G to M type members. Our final sample contains 288 sources with both $g$- and $r$-band lightcurves.

\subsection{LAMOST Spectroscopy}
\label{sec:data:lamost}

The Large Sky Area Multi-Object Fiber Spectroscopic Telescope (LAMOST), also called the Guoshoujing Telescope, is a quasi-meridian reflecting Schmidt telescope located at Xinglong Observatory Station in Hebei province, China. The telescope has an effective aperture of $\sim$4\;m and a field of view of 5$^{\circ}$ in diameter. The telescope is equipped with 16 spectrographs and 4000 fibers, each spectrograph has a resolving power of $R\approx1800$\footnote{The spectrographs have been upgraded to support median resolution observations with $R\approx7500$ since 2018 \citep{Liu2020arXiv200507210L}.}, and the wavelength coverage is 3700$-$9100\;\AA \citep{Cui2012RAA....12.1197C,Zhao2012RAA....12..723Z,Liu2015RAA....15.1089L,Luo2015RAA....15.1095L}.

Cross matching our working sample with the data release 9 of the LAMOST survey (LAMOST DR9\footnote{\url{http://www.lamost.org/dr9/}}), we obtain LAMOST spectra for 174 members in our working sample. There are 151 sources showing prominent H$\alpha$ emission lines in their LAMOST spectra. The accretion properties of these H$\alpha$ emitters are studied in Section~\ref{sec:accretion}.

\section{Target Properties}
\label{sec:star_disk_acc}

\subsection{Stellar Masses and Ages}
\label{sec:star:MassAge}

\begin{figure}[!t]
    \centering
    \includegraphics[width=\columnwidth]{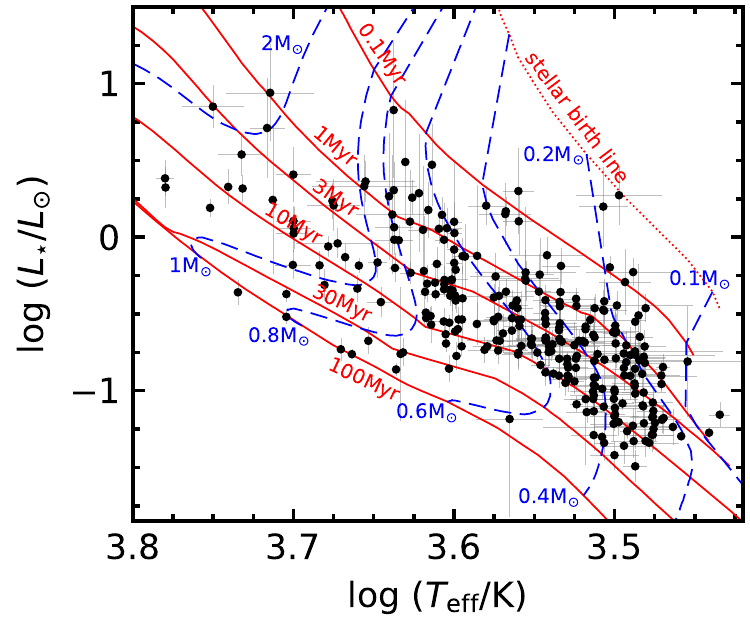}
    \caption{H-R diagram of the objects in our working sample. Overlaid are the isochrones (red solid lines) and mass tracks (blue dashed line) with solar metallicity from the PARSEC stellar model \citep{Bressan2012MNRAS.427..127B}, with their corresponding ages and masses indicated. The red dotted line is the stellar birth line.}
    \label{fig:HRD}
\end{figure}

\begin{figure}[!t]
    \centering
    \includegraphics[width=\columnwidth]{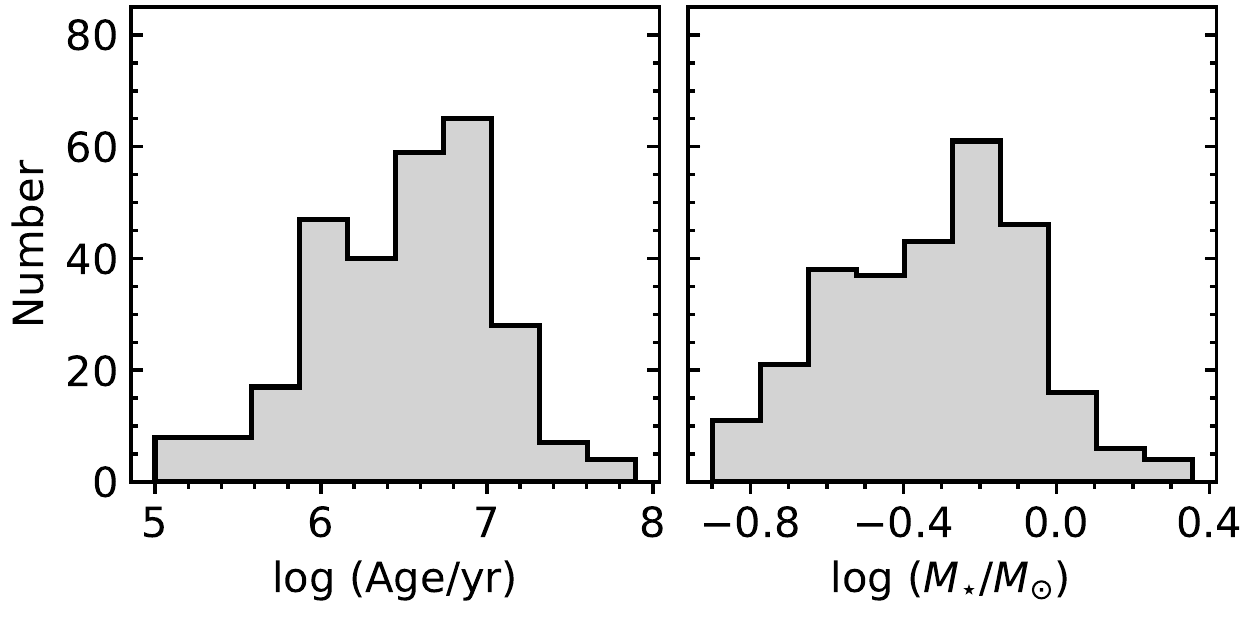}
    \caption{Histograms showing the distribution of stellar ages (left) and stellar masses (right) for the objects in our working sample. These values are estimated using the PARSEC stellar model \citep{Bressan2012MNRAS.427..127B} without correcting the contribution from spots.}
    \label{fig:hist_age_and_mass}
\end{figure}

Spectral types and extinction corrections have been provided for the full YSO sample \citep{Luhman2016ApJ...827...52L,Esplin2017AJ....154..134E,Luhman2020AJ....160...57L,Kounkel2019AJ....157..196K,Wang2022ApJ...936...23W}. We use the same methods as described in \citet{Wang2022ApJ...936...23W} to convert spectral types and observed $J$ magnitudes to effective temperatures and bolometric luminosities, respectively, and then to construct the Hertzsprung-Russell (H-R) diagram (Figure~\ref{fig:HRD}). Stellar masses and ages are estimated from their locations on the H-R diagram for individual sources using the PARSEC stellar model \citep{Bressan2012MNRAS.427..127B}. In Figure~\ref{fig:hist_age_and_mass}, we display the distributions of stellar ages and stellar masses. Though they span a large range of age, most objects in our working sample have ages between 1 to 10\;Myr, with median age of 3.8\;Myr. More than 90\% of the objects in our working sample are less massive than $1\;M_{\odot}$, and the median mass is $0.5\;M_{\odot}$.

\subsection{Disk Classification}
\label{sec:disk:class}

\begin{figure}[!t]
    \centering
    \includegraphics[width=\columnwidth]{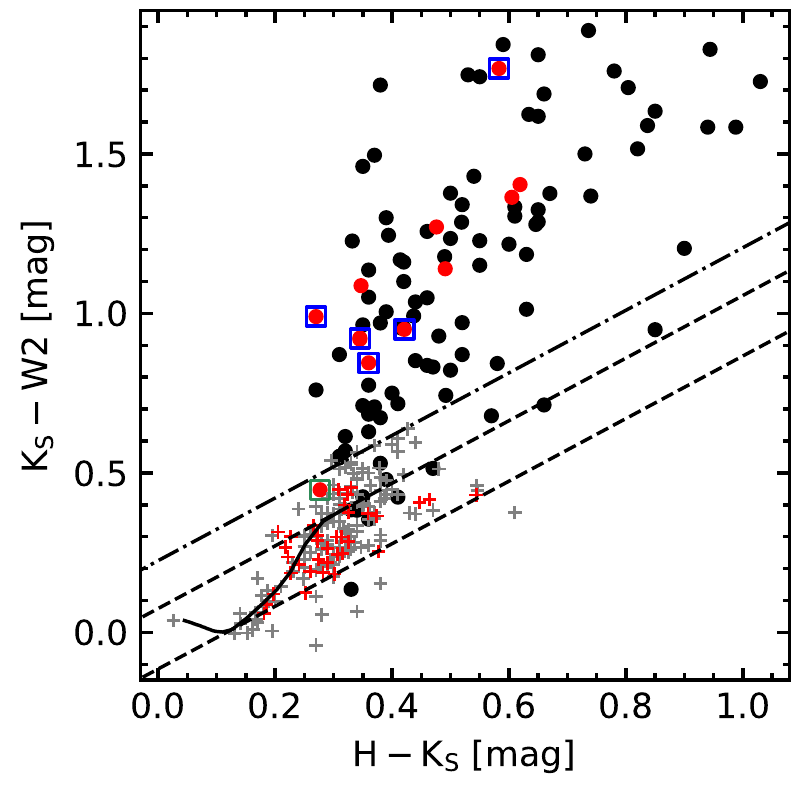}
    \caption{Infrared color-color plot for objects in our working sample. The solid circles and plus signs represent disked and diskless YSOs, respectively. Objects classified by us are highlighted as red. Additional blue and green squares mark sources that are classified as diskless in the literature. The solid curve is the locus of dwarfs from \citet{Pecaut2013ApJS..208....9P}, and the dashed lines correspond to the extinction law from \citet{Wang2019ApJ...877..116W}, enclosing the color space of dwarfs due to reddening. The dash-dotted line is the dividing line we use to separate disked YSOs from diskless ones (see Section~\ref{sec:disk:class} for detail).}
    \label{fig:CCD_KW2_HK}
\end{figure}

Most of the objects in the full YSO sample have disk classifications based mainly on \textit{Spitzer} photometry \citep{Dunham2015ApJS..220...11D,Luhman2016ApJ...827...52L,Kounkel2019AJ....157..196K}. We reclassified 6 of objects as disks based on their very red $K_{S}-W2$ colors (marked with blue squares in Figure~\ref{fig:CCD_KW2_HK}). One additional object is also reclassified as a disk based on its excess emission at $W4$ and $MP1$ bands (marked with green square in Figure~\ref{fig:CCD_KW2_HK}). Forty objects in our working sample with no disk classifications from the literature are classified here based on their locations on the $K_{S}-W2$ versus $H-K_{S}$ color-color diagram (Figure~\ref{fig:CCD_KW2_HK}). Objects with $K_{S}-W2$ colors redder than $0.98\times(H-K_{S})+0.22$ are classified as disks and bluer as diskless \citep{Wang2022ApJ...936...23W}. The dividing line separating disks from diskless YSOs is constructed as following. The locus of objects having disk classifications from literature are compared to the dwarf locus from \citet{Pecaut2013ApJS..208....9P} and reddening vector from \citet{Wang2019ApJ...877..116W}. Visually inspecting the color-color plot indicates that vertically shifting the upper border of the dwarf locus due to reddening redward of 0.15\;mag can separate disked YSOs from diskless ones fairly well. Five of these 40 objects are classified as disks, and the remaining as diskless ones. We also note several sources classified as disks in the literature are located at the diskless boundary in Figure~\ref{fig:CCD_KW2_HK}, because their infrared excess is seen only at wavelengths longer than $W2$.

Our working sample comprises of 109 disk objects and 179 diskless objects. The disk fraction of objects in our working sample is 38\%, slightly lower than that of the initial sample (46\%). This discrepancy is mainly due to that our working sample is constructed based on the ZTF photometry, which may be biased against low mass or embedded objects and stars with edge-on disks. We note that the disked and diskless objects in our working sample share similar mass ranges, and the majority of both samples are less massive than $1\;M_{\odot}$. The KS-test indicates that the two samples are indistinguishable in terms of spectral types ($p=7\%$).

\subsection{Accretion Properties}
\label{sec:accretion}

Accreting YSOs are generally characterized by strong and broad emission lines in their optical to near-infrared spectra \citep{Hartmann1994ApJ...426..669H,Muzerolle1998AJ....116.2965M,Muzerolle1998AJ....116..455M}. Correlations between emission line properties and accretion have been established both theoretically and observationally \citep[e.g.,][]{Muzerolle1998AJ....116.2965M,Natta2004A&A...424..603N,Fang2009A&A...504..461F}. H$\alpha$ is one of the strongest emission lines in classical T Tauri stars (CTTSs) and has been widely used as an indicator of accretion activity \citep{White2003ApJ...582.1109W,Muzerolle2003ApJ...592..266M,Natta2004A&A...424..603N,Fang2009A&A...504..461F,Fang2013ApJS..207....5F}.

In this section, we use the H$\alpha$ emission lines to study the accretion activities for a sub-sample of our working sample. We use the equivalent widths of H$\alpha$ emission lines ($\rm EW_{H\alpha}$) to distinguish between CTTSs and weak-line T Tauri stars (WTTSs) for the disk population. Since there are no unique $\rm EW_{H\alpha}$ value to distinguish all CTTSs from WTTSs, due to the ``contrast effect" \citep{Basri1995AJ....109..762B} and line optical depths \citep{Ingleby2011ApJ...743..105I}, we adopt the spectral type dependent thresholding values from \citet{Fang2009A&A...504..461F} to distinguish between CTTSs and WTTSs, that is an object is classified as CTTS if $\rm EW_{H\alpha}\ge3\;\AA$ for K0-K3 stars, $\rm EW_{H\alpha}\ge5\;\AA$ for K4 stars, $\rm EW_{H\alpha}\ge7\;\AA$ for K5-K7 stars, $\rm EW_{H\alpha}\ge9\;\AA$ for M0-M1 stars, $\rm EW_{H\alpha}\ge11\;\AA$ for M2 stars, $\rm EW_{H\alpha}\ge15\;\AA$ for M3-M4 stars, $\rm EW_{H\alpha}\ge18\;\AA$ for M5-M6 stars, $\rm EW_{H\alpha}\ge20\;\AA$ for M7-M8 stars. We further refine the classification by assigning the diskless objects as WTTSs. Forty sources are classified as CTTSs, and 134 sources are classified as WTTSs. Of the WTTSs, 86\% (116/134) are diskless objects.

\begin{figure}[!t]
    \centering
    \includegraphics[width=\columnwidth]{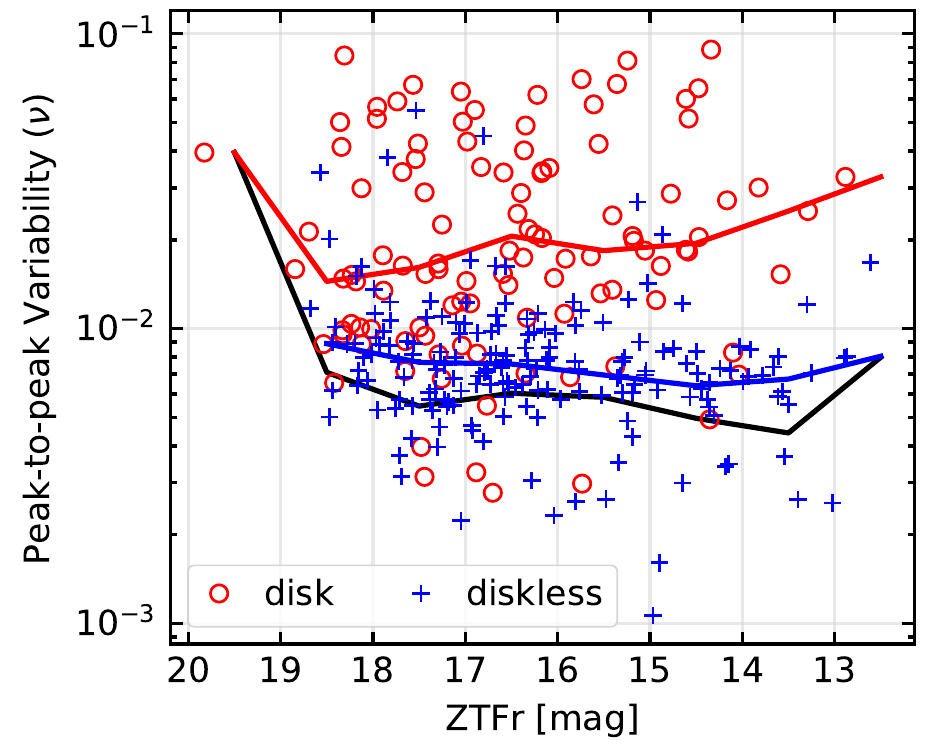}
    \caption{Normalized peak-to-peak variability metric $\nu$ as a function of mean $r$ magnitude for our working sample. Disked and diskless objects are indicated with red circles and blue pluses, respectively. The black solid line is the 15th percentile line, i.e., the boundary line we use to distinguish between variables and non-variables. The red and blue solid lines mark the median trends for disk and diskless populations, respectively.}
    \label{fig:ZTFr_vs_nu}
\end{figure}

\begin{figure*}[!t]
    \centering
    \includegraphics[width=0.96\textwidth]{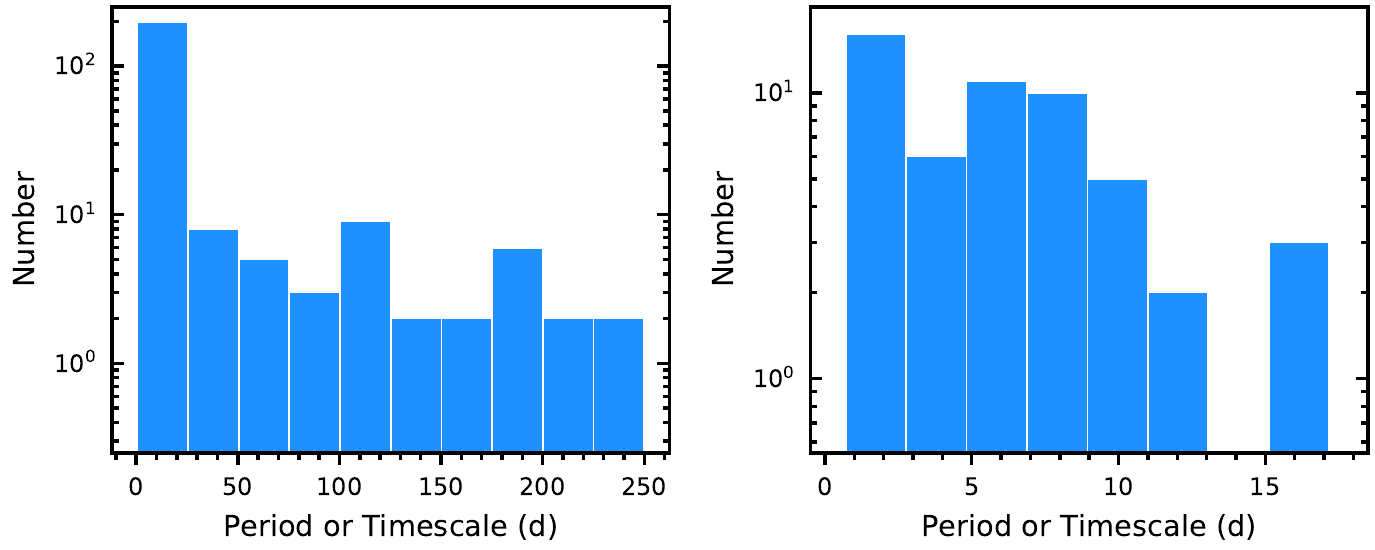}
    \caption{Histograms showing the distribution of periods or timescales of the periodogram peaks. Left panel shows all timescales, and right panel shows only sources classified as periodic variables.}
    \label{fig:hist_timescale}
\end{figure*}

\begin{figure}[!t]
    \centering
    \includegraphics[width=\columnwidth]{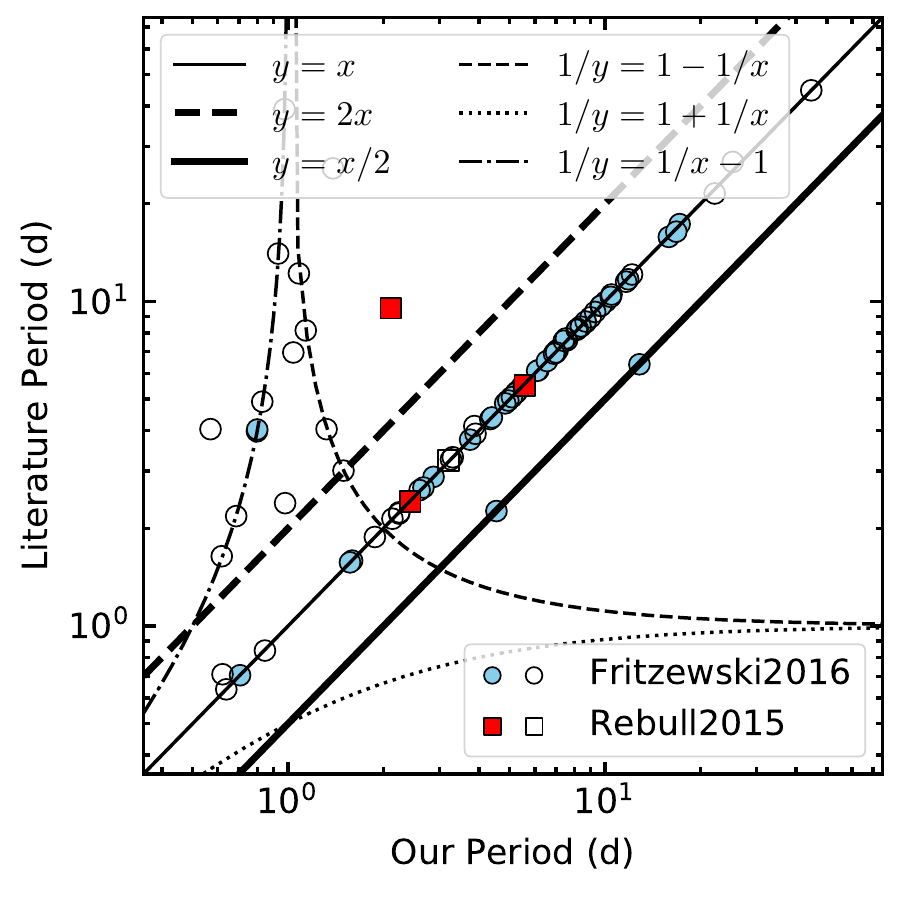}
    \caption{Comparison of periods between our work and previous literature for sources we classified as periodic (filled symbols) or quasiperiodic symmetry (open symbols). Circles are periods from \citet{Fritzewski2016MNRAS.462.2396F} and squares are periods from \citet{Rebull2015AJ....150..175R}. Different lines indicate the different harmonics and the alias with the $\sim$1\;day sampling as shown in the legend. The $x$ in the legend represents our periods and $y$ is literature periods.}
    \label{fig:PorT_vs_lit}
\end{figure}

\section{Lightcurve Analysis}
\label{sec:lc_analysis}

Most objects in our working sample have 200$-$500 observations during the $\sim$4 years ZTF data stream. Since the $r$-band photometry is much better than the $g$-band and the cadence is generally higher in the $r$-band than in the $g$-band, our analysis of the lightcurves in this section is mainly based on the $r$-band photometry. The $g$-band photometry is only used in analyzing the CMD pattern and when mentioned specifically.

\subsection{Variability Search}
\label{sec:var_search}

We use the normalized peak-to-peak variability metric \citep{Sokolovsky2017MNRAS.464..274S}
\begin{equation}
    \nu = \dfrac{(m_{i}-\sigma_{i})_{\rm max}-(m_{i}+\sigma_{i})_{\rm min}}
                {(m_{i}-\sigma_{i})_{\rm max}+(m_{i}+\sigma_{i})_{\rm min}},
\end{equation}
to measure variability amplitude for objects in our working sample, where $m_{i}$ is a magnitude measurement and $\sigma_{i}$ is the corresponding measurement uncertainty. The maximum and the minimum are determined from the full light curve. The normalized peak-to-peak variability is a sensitive variability indicator \citep{Sokolovsky2017MNRAS.464..274S} since we have removed potential outliers from each light curve (see Section~\ref{sec:data:ZTF}).

Following \citetalias{Hillenbrand2022AJ....163..263H}, we consider an object variable if its $\nu$ metric is greater than the 15th percentile of $\nu$ as a function of mean $r$ magnitude (Figure~\ref{fig:ZTFr_vs_nu}). As pointed out in \citetalias{Hillenbrand2022AJ....163..263H}, although this is not a rigorously justified cutoff, this cut ensures that we select the fractionally larger amplitude objects as variables at each brightness level. We select 238 (83\% of our working sample) objects as variables based on the $\nu$ metric.

92\% of the disk population and 77\% of the diskless population are variables. The variability fraction of the disk population is much higher than that of the diskless population, as can be more evidently seen in Figure~\ref{fig:ZTFr_vs_nu}. In addition, the $\nu$ metrics are typically 1-3 times larger for disk population than for diskless population (as indicated with red and blue solid lines in Figure~\ref{fig:ZTFr_vs_nu}).

The properties of the 238 variables are listed in Table~\ref{tab:variables}. The variability amplitude listed in the table is calculated as the difference between the 5th and 95th percentile magnitudes.

\begin{figure*}[!t]
    \centering
    \includegraphics[width=0.9\textwidth]{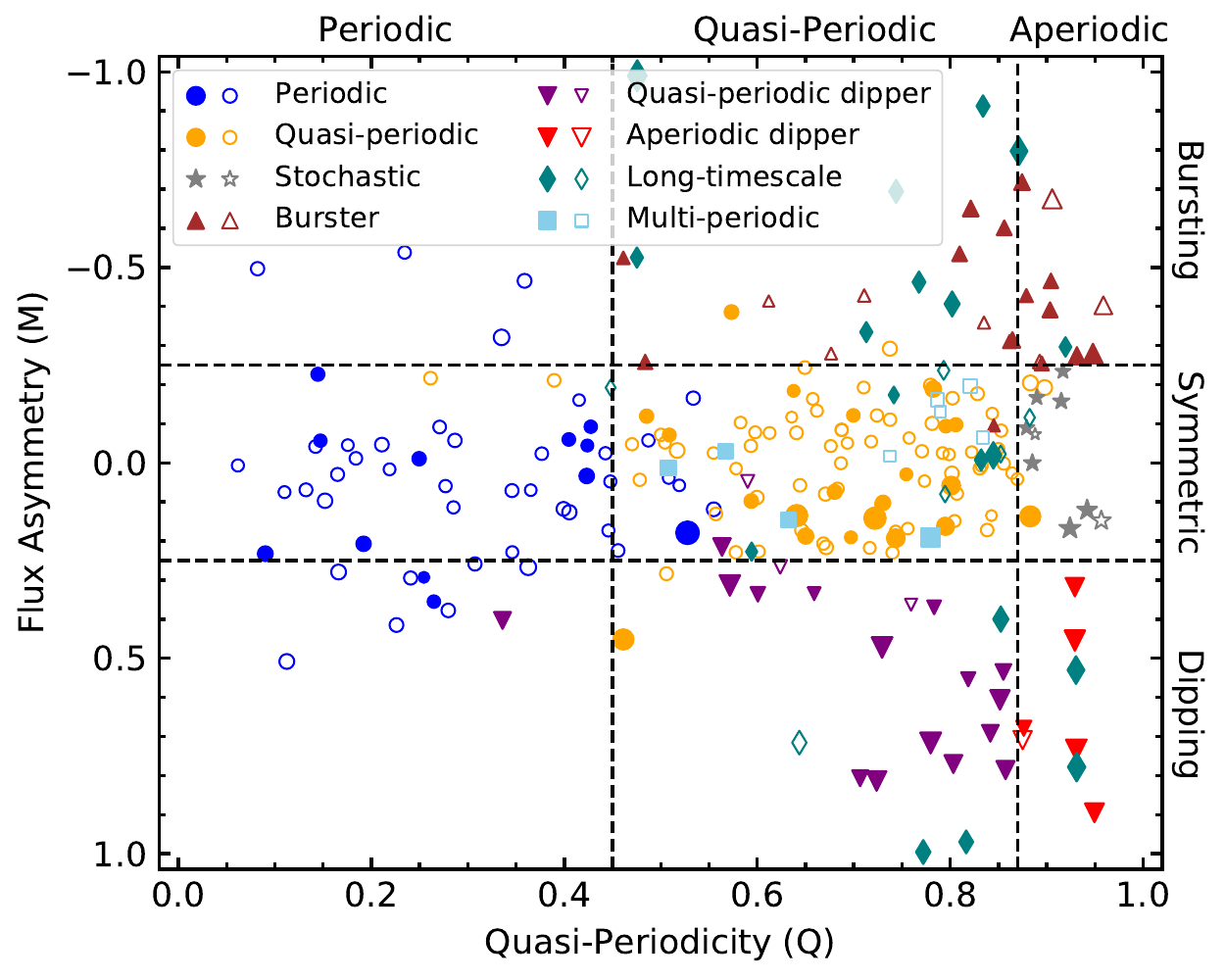}
    \caption{Quasiperiodicity ($Q$) vs. flux asymmetry ($M$) for our sample of variables, color coded by lightcurve morphological types. The sizes of the points are proportional to the square root of the normalized peak-to-peak variability metric $\nu$. Disked objects are indicated with solid symbols and diskless objects are marked with open symbols. Note that unclassifiable variables are not included in this plot, since they fall off the viewing range of the plot.}
    \label{fig:Q_vs_M}
\end{figure*}

\subsection{Period Search}
\label{sec:period_search}

 We use the Astropy \citep{Astropy-Collaboration2013A&A...558A..33A,Astropy-Collaboration2018AJ....156..123A} implementation of the Lomb-Scargle periodogram \citep{Lomb1976Ap&SS..39..447L,Scargle1982ApJ...263..835S,VanderPlas2015ApJ...812...18V,VanderPlas2018ApJS..236...16V} to compute the periodogram for each variable and search for periods between a minimum period of 0.5\;days and a maximum period of 250\;days. Periods between 0.5$-$0.51, 0.98$-$1.02, 1.96$-$2.04, and 26$-$30 days are flagged for further analysis to avoid the most common aliases associated with the solar and sidereal days, as well as the lunar cycle \citep{Rodriguez2017ApJ...848...97R,Ansdell2018MNRAS.473.1231A,Hillenbrand2022AJ....163..263H}. We also reject periods with half or double multiples that fail at least one of the aforementioned alias checks to account for additional potential aliases. In most cases, the period is searched using the $r$-band lightcurves, except for cases where the $g$-band lightcurve is better phased.

We then use the \texttt{find\_peaks}\footnote{\url{https://docs.scipy.org/doc/scipy/reference/generated/scipy.signal.find_peaks.html}} function from the PYTHON package \texttt{scipy} \citep{Virtanen2020NatMe..17..261V} to search for power peaks in the periodogram. Only peaks higher than the 1\% false alarm level\footnote{Since the true probability distribution for the largest peak cannot be determined analytically, we estimate the false alarm probability approximately using the approach of \citet{Baluev2008MNRAS.385.1279B}.} are considered significant, and are retained for further analysis. If there are no peaks more significant than the 1\% false alarm probability, we take the period corresponding to the maximum power of the periodogram as estimates of the variability timescales. In most cases the peak with the maximum power is adopted as the real period, but there are cases where the peak with slightly lower power is adopted as the real for better interpreting the beat patterns and improving the phase dispersion minimization (i.e., with smaller $Q$ values determined in the next section).

For sources with multiple peaks in the periodogram, the frequencies of the peaks are checked for possible alias and beats following the proposals in \citet{VanderPlas2018ApJS..236...16V}. The source is labeled as multiperiodic if there are peaks with frequencies that are not alias or beats of the adopted real period. Based on our visual inspection, only the first four highest peaks are checked for sources with more than four peaks in their periodograms.

The distribution of the periods or timescales are displayed in Figure~\ref{fig:hist_timescale}. Our variable sample has a relatively flat distribution, extending to more than 200\;days. For the periodic objects, our sample is double peaked with a tail toward longer periods. The two peaks at around 1.5\;days and one week are consistent with the populations of fast and slow rotators, respectively \citep{Gallet2013A&A...556A..36G}.

We compare our periods with that from literature \citep[e.g.,][]{Fritzewski2016MNRAS.462.2396F,Rebull2015AJ....150..175R} in Figure~\ref{fig:PorT_vs_lit}. In most cases, our periods are consistent with that from literature. The largest discrepancy is found for objects we classified as quasiperiodic, for which we found periods that are aliases of the literature periods with the 1\;day sampling.

\begin{table*}
\centering%
\renewcommand{\arraystretch}{1.5}\setlength{\tabcolsep}{3ex}
\caption{Criteria used to classify the lightcurves into different variability categories.}\label{tab:lcType_criteria}
\begin{tabular}{lll}\toprule
$Q$            & $M$            & Variability Type              \\\midrule
$Q<0.45$       & $-0.25<M<0.25$ & Periodic [P]                  \\
$0.45<Q<0.87$  & $-0.25<M<0.25$ & Quasiperiodic Symmetric [QPS] \\
$0.45<Q<0.87$  & $M>0.25$       & Quasiperiodic Dipper [QPD]    \\
$Q>0.87$       & $M>0.25$       & Aperiodic Dipper [APD]        \\
$\cdots\cdots$ & $M<-0.25$      & Burster [B]                   \\
$Q>0.87$       & $-0.25<M<0.25$ & Stochastic [S]                \\
$Q>1$ or $Q<0$ & $\cdots\cdots$ & Unclassifiable [U]            \\\bottomrule
\end{tabular}
\end{table*}

\subsection{Light-curve Classification}
\label{sec:lc_class}

We classify our light-curves into different types based on two statistics quantifying quasiperiodicity ($Q$) and flux asymmetry ($M$), first developed by \citet{Cody2014AJ....147...82C} based on regularly sampled time series from space-based platforms and further refined by \citet{Cody2018AJ....156...71C}. The $Q$ and $M$ metrics are defined as follows, 
\begin{equation}
\begin{cases}
Q=\dfrac{\sigma_{\rm resd}^{2}-\sigma_{\rm phot}^{2}}{\sigma_{m}^{2}-\sigma_{\rm phot}^{2}} \\[3ex]
M=\dfrac{\left<m_{10\%}\right>-m_{\rm med}}{\sigma_{m}}
\end{cases},
\end{equation}
where $\sigma_{m}$ is the scatter of the original light-curve, $\sigma_{\rm resd}$ is the scatter of the residual light-curve after subtracting the smoothed dominant periodic signal, $\sigma_{\rm phot}$ is taken as the mean photometric error of all observations in an object's light-curve, scaled by a factor of 1.25 to account for an initial compression of $Q$ values \citepalias{Hillenbrand2022AJ....163..263H}, $m_{\rm med}$ is the median magnitude of the light-curve, and $\left<m_{10\%}\right>$ is the mean magnitude of the top and bottom 10\% measurements. The reader is referred to \citetalias{Hillenbrand2022AJ....163..263H} for more details on the $Q$ and $M$ metrics. In most cases, the $Q$ and $M$ metrics are calculated for each variable using the $r$-band lightcurve data. For several cases, the period is determined for the $g$-band lightcurve only, and the $g$-band lightcurve is used to determine these statistics. The variables in our working sample are classified into 9 categories, based on their locations in the $Q-M$ plane (Figure~\ref{fig:Q_vs_M}) and additional visual inspection.

Since the time series data analyzed here is from the same instrument as in \citetalias{Hillenbrand2022AJ....163..263H}, we adopt the same boundary values of $Q$ and $M$ metrics as \citetalias{Hillenbrand2022AJ....163..263H}. The boundary values used to classify the lightcurves into different variability categories are summarized and listed in Table~\ref{tab:lcType_criteria}. In addition to the seven categories listed in Table~\ref{tab:lcType_criteria}, we classify objects with variability timescales larger than 100\;day as long timescale (L) variables, and objects having more than one periods are classified as multi-periodic (MP) variables. All of the lightcurves and corresponding periodograms are visually inspected. In most cases, the $Q$ and $M$ classifications are consistent with our visual inspection. For 40 cases (labeled in Table~\ref{tab:variables}), we adjust their type to favor our visual classification. We note that the original classification scheme does not involve the $Q-M$ plane with $0<Q<0.45$ and $M>0.25$. About 10 of these cases have $(Q,\;M)$ values in this region, and these sources are classified as periodic variables based on their well phased lightcurves. Other adjustments mainly occur around the boundaries, and the most common adjustment is from quasiperiodic symmetric category to periodic type for seven objects due to their well-phased lightcurves and their quite clean periodograms. These adjustment does not affect our statistics significantly. We should mention that both the photometric precision and the cadence will affect the measurement of the $Q$ metric, and thus the classification of the lightcurve morphology. The readers are referred to \citetalias{Hillenbrand2022AJ....163..263H} for a discussion on these effects (see their Appendix C).

\begin{figure}[!t]
    \centering
    \includegraphics[width=\columnwidth]{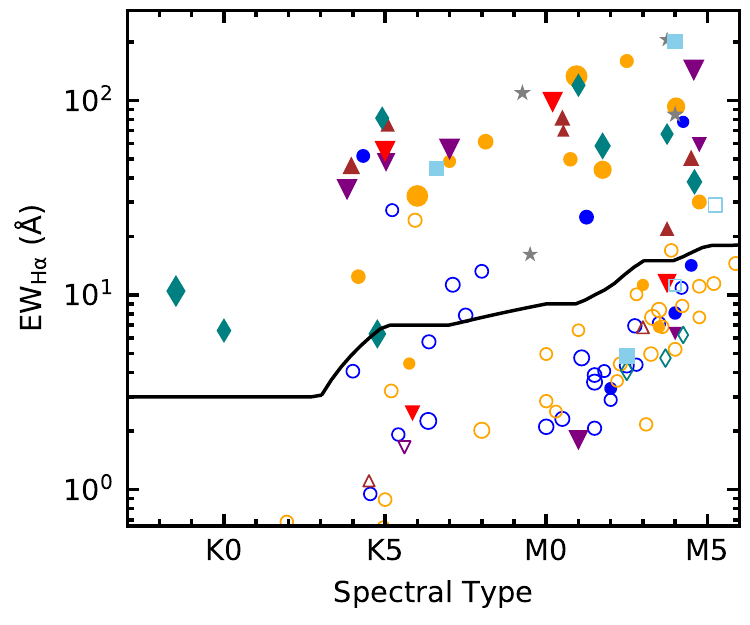}
    \caption{Equivalent width of H$\alpha$ emission line as a function of spectral type. The symbols are the same as in Figure~\ref{fig:Q_vs_M}. The solid line is the dividing line separating CTTSs from WTTSs \citep{Fang2009A&A...504..461F}.}
    \label{fig:EW_vs_SPT}
\end{figure}

\begin{table*}
\centering%
\renewcommand{\arraystretch}{1.5}\setlength{\tabcolsep}{1.5ex}
\caption{Distribution of lightcurve morphological category for objects in our variable sample.}\label{tab:lcType}
\begin{tabular}{lcccccc}\toprule
\multirow{2}*{Variability Type} & \multicolumn{3}{c}{Numbers}  & \multicolumn{3}{c}{Numbers} \\\cmidrule(lr){2-4}\cmidrule(lr){5-7}
                                & disk & diskless & total           & CTTS & WTTS & total\\\midrule
                  Periodic [P]  &    12  &    41  &    53  &     3  &    28  &    31\\
 Quasiperiodic Symmetric [QPS]  &    22  &    71  &    93  &    10  &    58  &    68\\
           Multi-periodic [MP]  &     4  &     5  &     9  &     2  &     4  &     6\\
                Stochastic [S]  &     7  &     2  &     9  &     4  &     1  &     5\\
                   Burster [B]  &    15  &     7  &    22  &     6  &     2  &     8\\
    Quasiperiodic Dipper [QPD]  &    16  &     3  &    19  &     5  &     6  &    11\\
        Aperiodic Dipper [APD]  &     5  &     1  &     6  &     2  &     2  &     4\\
            Long Timescale [L]  &    18  &     6  &    24  &     7  &     5  &    12\\
            Unclassifiable [U]  &     0  &     2  &     2  &     0  &     1  &     1\\\midrule
                  Total number  &   100  &   138  &   238  &    39  &   107  &   146\\\bottomrule
\end{tabular}
\end{table*}

\begin{figure*}[!t]
    \centering
    \includegraphics[width=\textwidth]{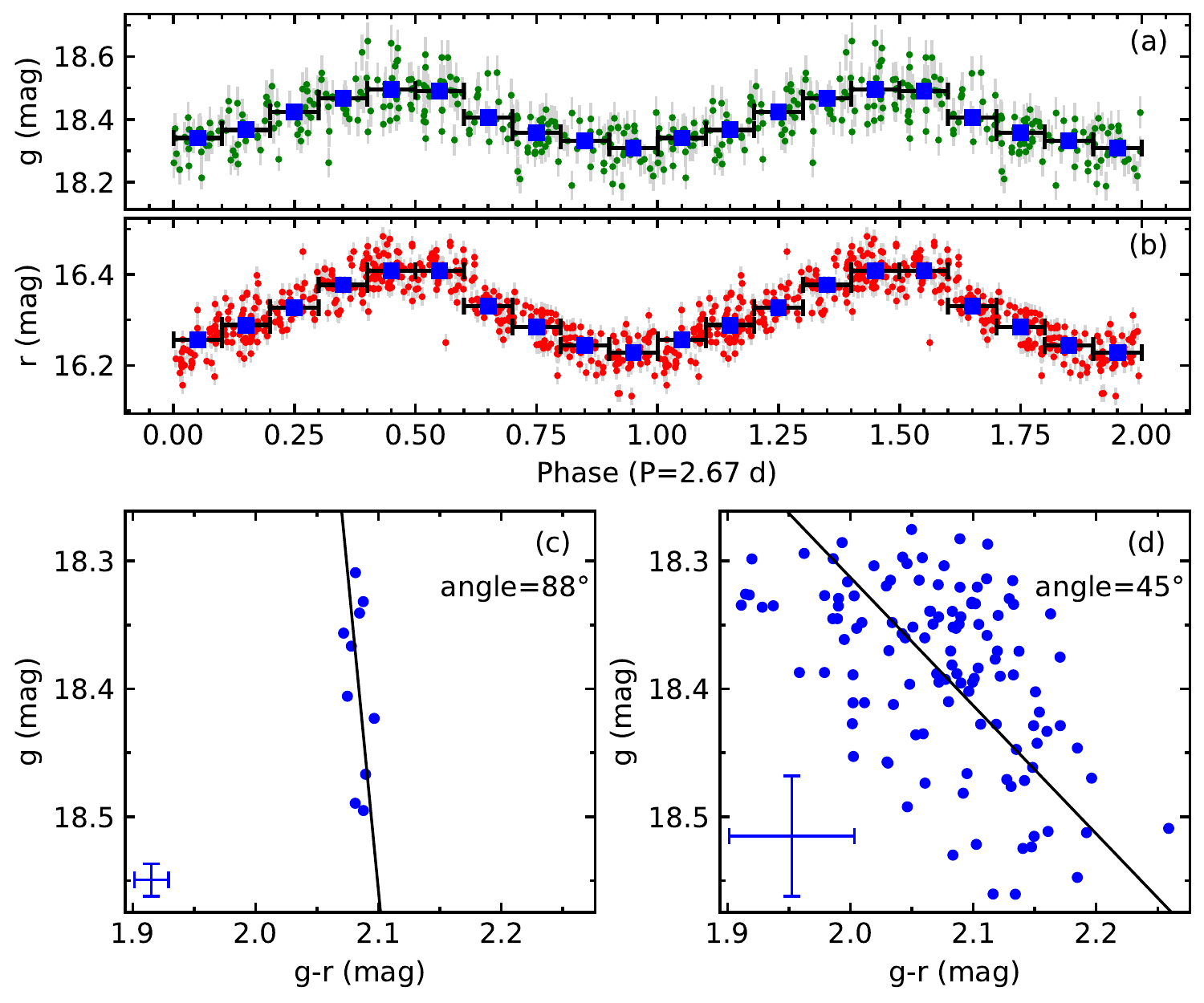}
    \caption{An example displays constructing the phase series CMD for the source 2MASS J03442812+3216002. Panel (a): The $g$-band phase-folded lightcurve is displayed as small points with error bars. The blue squares show the median magnitudes in individual phase bins, and the horizontal error bars are the corresponding phase bins. The standard deviations of the median magnitudes are smaller than the symbol size. Panel (b): Similar as panel (a), but for the $r$-band data. Panel (c): The phase series CMD is displayed as blue points. The black line is the straight line from the orthogonal regression. The error bar in the lower left displays the typical uncertainty of the data points. The corresponding CMD slope angle is labeled. Panel (d): Similar as panel (c), but displaying the time series CMD for comparison.}
    \label{fig:Method2}
\end{figure*}

\begin{figure*}[!t]
    \centering
    \includegraphics[width=\textwidth]{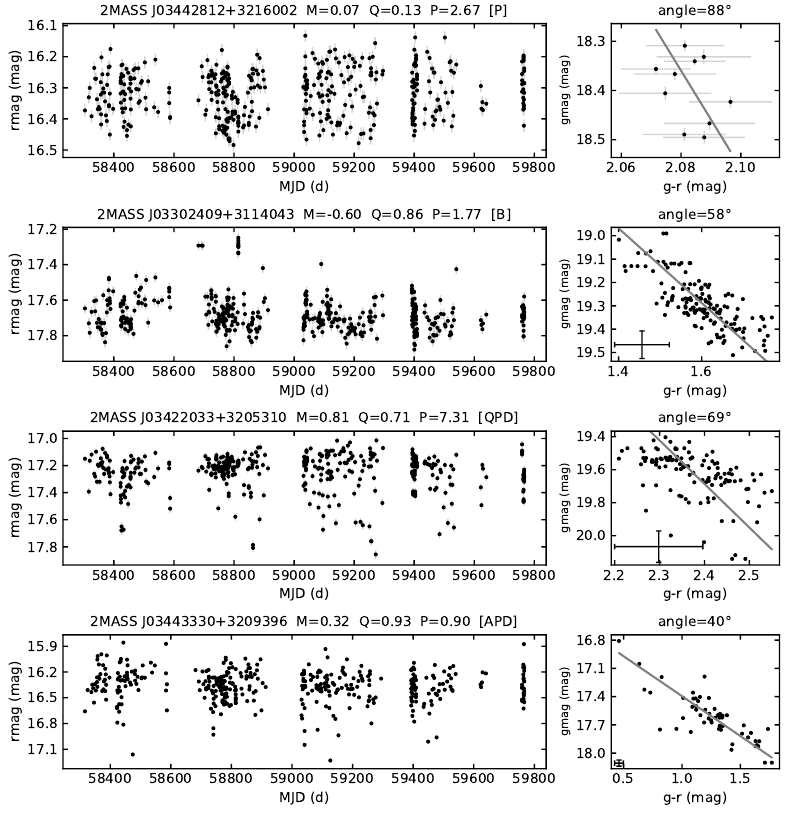}
    \caption{Examples showing the lightcurves (left panels) and corresponding CMDs (right panels) for periodic, burster, quasiperiodic dipper and aperiodic dipper categories. In the left panels, the source name, the flux asymmetry ($M$), the quasiperiodicity ($Q$), the period/timescale and the lightcurve classification are labeled for each source. In the right panels, the gray lines are the straight line from the orthogonal regression, and the corresponding CMD slope angles are labeled.}
    \label{fig:example1}
\end{figure*}

\begin{figure*}[!t]
    \centering
    \includegraphics[width=\textwidth]{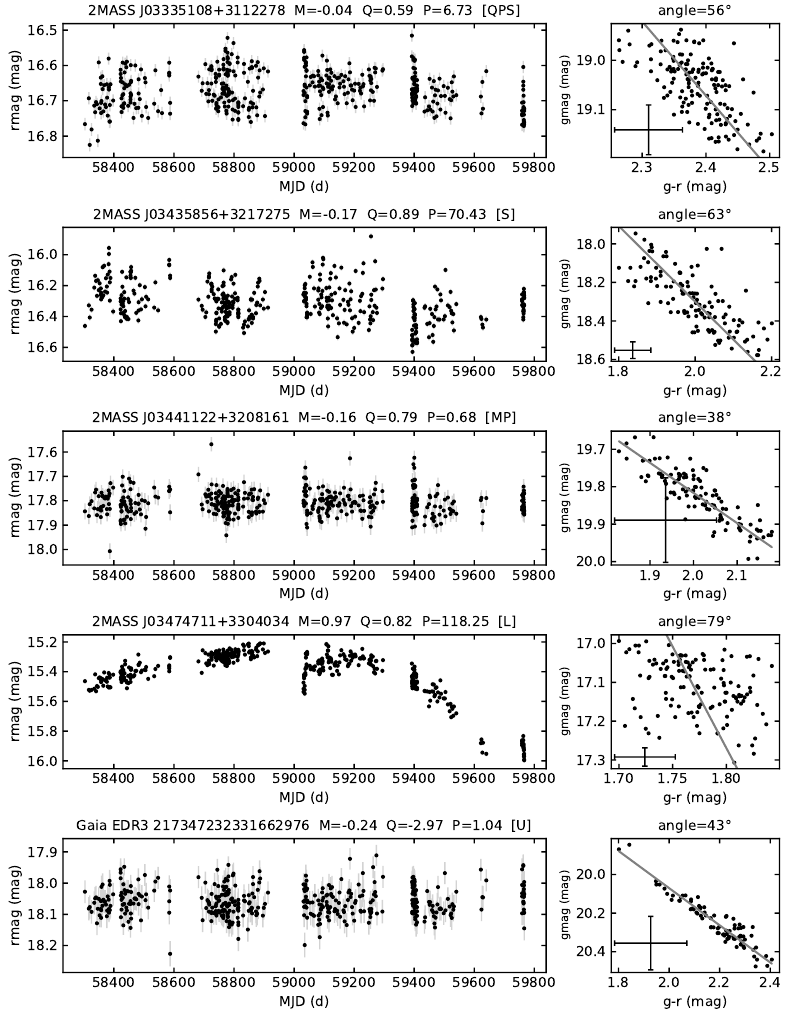}
    \caption{Same as Figure~\ref{fig:example1}, but for quasiperiodic symmetric, stochastic, multi-periodic, long timescale and unclassifiable categories.}
    \label{fig:example2}
\end{figure*}

\begin{figure*}[!t]
    \centering
    \includegraphics[width=\textwidth]{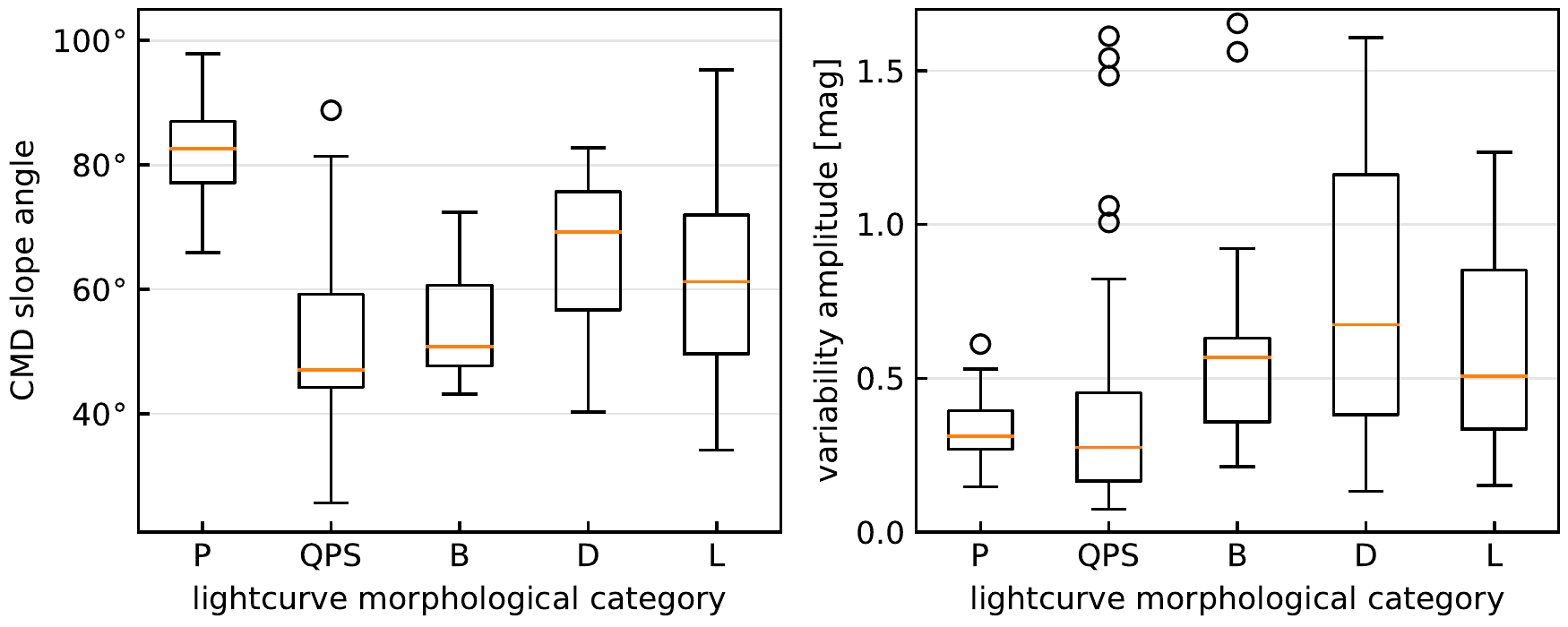}
    \caption{Boxplots showing the distributions of CMD slope angle (left) and variability amplitude (right) for different lightcurve morphological categories. Multi-periodic and unclassifiable categories are not included here.}
    \label{fig:boxplot_for_angamp}
\end{figure*}

The numbers of different lightcurve morphologies are listed in Table~\ref{tab:lcType}. The dominant lightcurve morphology in our variable sample is quasiperiodic symmetric. For the disk population, dipper (both quasiperiodic dippers and aperiodic dippers), burster, stochastic and long timescale categories are also common.  We performed two-dimensional Kolmogorov-Smirnov (KS) test \citep{Peacock1983MNRAS.202..615P}\footnote{We used the 2D KS-test PYTHON implementation \texttt{ndtest} available at \url{https://github.com/syrte/ndtest}.} to compare the 2D parameter space ($Q$, $M$) for disk and diskless populations. We found $p$-value of $2\times10^{-4}$ corresponding to the null-hypothesis that disk and diskless populations occupy the same ($Q$, $M$) space. The disk population is dominated by asymmetric and non-repeatable lightcurves, while the diskless population is dominated by symmetric and repeatable lightcurves. There are eleven diskless objects classified as bursters or dippers. Visually inspecting their lightcurves indicates that they have variability amplitudes comparable to the measurement uncertainties or their $M$ metrics are affected by several photometric measurements, which makes their classifications unreliable. As pointed out in Section~\ref{sec:disk:class}, the disk and diskless populations in our working sample share similar mass ranges and spectral type distributions. In addition, we don't find any trends of variability properties as functions of stellar masses or spectral types. Considering these issues together, the differences of the variability patterns of the two populations are dominated by the presence or absence of disks.

The $\rm EW_{H\alpha}$ value are displayed as a function of spectral type and lightcurve morphological category in Figure~\ref{fig:EW_vs_SPT} for a sub-sample of 146 variables having LAMOST spectra. In the figure, nearly all sources in the burster (brown upward triangles) and stochastic (gray star symbols) categories are CTTSs. Half of the dippers and long timescale variables are also CTTSs. More than 85\% (90/105) of variables classified as P, QPS, or MP are WTTSs. The numbers of different lightcurve types in this sub-sample are listed in Table~\ref{tab:lcType}. These are consistent with the results on the different properties of the disked or diskless variables discussed above.

\subsection{CMD Analysis}
\label{sec:cmd_analysis}

There are many different physical mechanisms that can drive the photometric variability observed in young stars (see \citetalias{Hillenbrand2022AJ....163..263H} for a summary of the mechanisms related to different lightcurve morphologies). Color time series data is a powerful tool in distinguishing these physical mechanisms. Our working sample is constructed to have both $g$ and $r$-band lightcurves, so it is possible for us to analyse $g-r$ color times series data, besides $g$ or $r$-band lightcurves.

Since the ZTF $g$ and $r$-band observations are not simultaneous, with time steps of a few hours up to a few days, we construct the CMDs with the following two methods.

The first method (Method~1) is the same as that in \citetalias{Hillenbrand2022AJ....163..263H}. For each source, the $r$-band lightcurve is trimmed into the corresponding $g$-band time span. For each time point in the trimmed $r$-band lightcurve, we search the $g$-band lightcurve for paired observations with one just before this time and another after this time, we then linearly interpolate the paired $g$-band observations to this time point and to estimate the $g-r$ colors, if the time interval of the paired $g$-band observations is less than 3\;days. The errors of the interpolated $g$ magnitudes and $g-r$ colors are estimated using the PYTHON package \texttt{uncertainties} \citep{Lebigot2010uncertainties}.

The second method (Method~2) we developed to construct the CMD is based on the phase-folded lightcurves and applied to only periodic objects in our sample. For periodic objects, both the $g$ and $r$-band lightcurves are phase-folded at the adopted periods. The median magnitudes, the standard deviations and the number counts are determined for 10 evenly spaced phase bins, for both the $g$ and $r$-band lightcurves, for each source. The errors of the magnitudes corresponding to each phase bin is estimated as the standard deviation divided by the square root of the number count, of that phase bin. The $g-r$ colors are estimated at the same phase, with the errors estimated using the \texttt{uncertainties} package as well. The CMD constructed this way will be designated the phase series CMD in the remaining of the paper. An example of constructing the phases series CMD is displayed in Figure~\ref{fig:Method2}.

\begin{figure*}[!t]
    \centering
    \includegraphics[width=\textwidth]{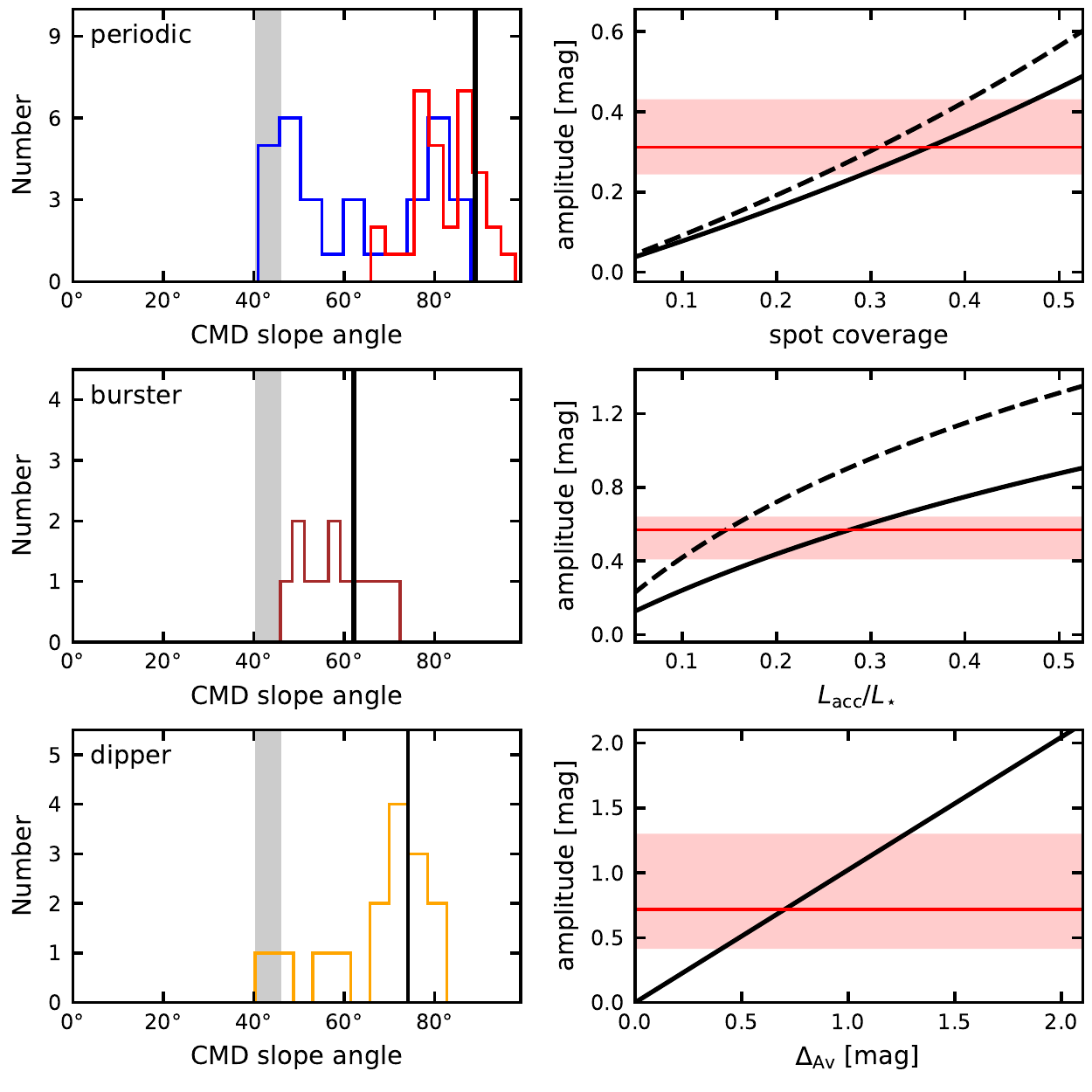}
    \caption{Left panels: histograms showing the distributions of CMD slope angles for periodic, burster and dipper variables from top to bottom, as labeled in the corresponding panels. The gray shaded area display the $1\sigma$ range of the slope angles for non-variables in our sample (see the discussion in Appendix~\ref{app:noise}). The vertical line in each panel corresponds to the angles due to changes in spot coverage, variable accretion and variable extinction, respectively, from top to bottom. The vertical lines in the left top and middle panels are calculated for $T_{\rm eff}=4000\;\rm K$. The blue and red histograms in the left top panel represent the angles determined using Method~1 and Method~2, respectively. Right panels: the variability amplitudes observed are compared to that due to changes of spot coverage, variable extinction and variable accretion for periodic, burster and dipper variables, respectively, from top to bottom. The red horizontal line and the shaded area in each panel show the median value and the $1\sigma$ range of the amplitude for the corresponding sample, respectively. The solid and dashed lines in the right top and middle panels correspond to calculations for $T_{\rm eff}=4000\;\rm K$ and 3500\;K, respectively.}
    \label{fig:CMDpattern}
\end{figure*}

For each source in the variable sample, we fit a straight line to its time/phase series CMD using an orthogonal distance regression in the PYTHON package \texttt{scipy.odr} \citep{Virtanen2020NatMe..17..261V}, following \citep{Poppenhaeger2015AJ....150..118P,Hillenbrand2022AJ....163..263H}. This method is chosen to account for the significant and partially correlated errors in both axes. For the CMD constructed using the first method, only the central 95\% of the CMD spans in $g$ and $g-r$ are included in the regression, to alleviate the effect of outliers.

The slope angles are defined as the inverse tangent of the best-fitting slopes in the $g$ versus $g-r$ CMDs. The slope angles are expressed in degrees and span from $0^{\circ}$ (corresponding to color changes with no associated $g$-band variability) to $90^{\circ}$ (corresponding to $g$-band variability with no color changes). To estimate the errors on the best-fitting slope angles, the CMD points are perturbed according to the errors in $g$ magnitudes and $g-r$ colors assuming Gaussian errors (this is similar to the perturbation method described in \citet{Curran2014arXiv1411.3816C}) and the orthogonal distance regression is performed on the perturbed points. This procedure is repeated 1000 times for each source and the errors on the angles are estimated as the standard deviation of the 1000 realizations. Only sources with errors less than $10^{\circ}$ are considered for further analysis. This error estimation is slightly different from that in \citetalias{Hillenbrand2022AJ....163..263H}, where the errors are estimated using a bootstrap technique. Several examples of the constructed CMDs and corresponding lightcurves are displayed in Figure~\ref{fig:example1} and \ref{fig:example2}.

In Figure~\ref{fig:boxplot_for_angamp}, we display the distribution of CMD slope angle and variability amplitude for different lightcurve morphological categories. Strictly periodic objects show the largest angles, and bursters have much flatter slopes than dippers. Periodic and quasiperiodic sources have the lowest variability amplitudes. In the next section, we demonstrate that the angles of periodic variables are consistent with the spot model, the angles of bursters are consistent with the accretion model, and the angles of dippers are consistent with variable extinction. We also note that stochastic variables have angles in consistent with variable accretion, which may indicate that these sources are ongoing accretion activity as well \citep[as demonstrated in][]{Stauffer2016AJ....151...60S}. All but one of the stochastic variables having LAMOST spectra are accreting stars.

\section{Discussion}
\label{sec:CMDpattern_model}

In Section~\ref{sec:lc_class} we classified our variables into different categories and, in Section~\ref{sec:cmd_analysis} we analysed the CMD patterns using orthogonal distance regression technique on $g-r$ versus $g$ CMD. In this section, we will discuss the different CMD patterns of periodic, burster and dipper categories, and relate them with specific mechanisms.

There are 53 ($\sim$21\% of the variable sample) objects classified as strictly periodic variables, with CMD slope angles determined for 32 of them using Method 1 and Method 2. As shown in the upper left panel of Figure~\ref{fig:CMDpattern}, we obtain much larger angles using Method 2 than  using Method 1. These periodic variables are generally explained as stellar rotation modulated by star spots on the stellar surface. We model the CMD pattern for a star with $T_{\rm eff}=4000\;\rm K$ and a cool spot 500\;K cooler than the effective temperature on the stellar surface. The CMD trend arising from the cool spot is nearly vertical, corresponding to colorless variability with associated changes in $g$ magnitudes and the corresponding slope angle is essentially 90$^{\circ}$. \citet{Gully-Santiago2017ApJ...836..200G} also found nearly colorless changes in $B-V$ or $V-R$ colors with associated brightness changes in $V$-band. Changing the effective temperatures and the temperature contrasts does not alter the angles significantly. 

Comparing the slope angles we determined with this simple model, we find that the angles from  Method~2 are more consistent than from Method~1 with the cool spot model. One possible reason could be that the interpolation of the $g$-band observations to the $r$-band observing time in Method~1 introduce additional uncertainties, especially for fast rotators. We do note a trend of decreasing differences between the two methods with increasing period. The effect of noise on the CMD pattern is discussed in Appendix~\ref{app:noise}. We also compared the variability amplitude in $g$-band with the cool spot models (upper right panel of Figure~\ref{fig:CMDpattern}). The comparison indicates that the typical changes in spot coverage of our periodic variables is in the range 30\% to 40\%, and this range is consistent with the cool spot coverage in \citet{Cao2022MNRAS.517.2165C,Herbert2023MNRAS.520.5433H}. For our analysis, we model a single spot on the stellar surface, and the reader is referred to \citet{Guo2018ApJ...868..143G} for a discussion about multiple spots configuration. In addition, the spot coverage estimated from lightcurve amplitude alone may be underestimated \citep{Rackham2018ApJ...853..122R}.

Among the sample of variables with disks, 15 ($\sim$15\% of the sample) are bursters, and 12 of the 15 have CMD slope. Burster variables are thought to arise from discrete accretion shocks. We model the CMD pattern for a star with $T_{\rm eff}=4000\;\rm K$, and adopt the accretion spectrum from \citet{Manara2013A&A...558A.114M} with  the same model parameters as in \citet{Flaischlen2022A&A...666A..55F}, i.e., the electron temperature $T_{\rm slab}=11000\;\rm K$, the electron density $n_{\rm e}=10^{15}\;\rm cm^{-3}$, and the optical depth at $300\;\rm nm$ $\tau_{300}=5.0$. The CMD slope angles due accretion variation is around $62^{\circ}$, in consistence with the bursters in our sample (middle left panel of Figure~\ref{fig:CMDpattern}). We also compared the variability amplitudes in $g$-band with the accretion model (middle right panel of Figure~\ref{fig:CMDpattern}), and the comparison indicate that the bursters in our sample have changes of $L_{\rm acc}/L_{\star}$ in the range 0.1 to 0.3, these values are larger than the typical value of 0.11 in \citet{Flaischlen2022A&A...666A..55F}, but within 1$\sigma$ range of that work.

Among the sample of variables having disks, there are 21 ($\sim$21\% of the sample) dippers, with CMD slope angles determined for 15 of them. The slope angles of these dippers are $\sim$74.1$^{\circ}$, consistent with that expected from interstellar extinction according to the extinction law from \citet{Schlafly2011ApJ...737..103S} with total to selective extinction ratio of $R_{V}=3.1$. Comparing the variability amplitudes in $g$-band with that due to extinction changes (bottom right panel of Figure~\ref{fig:CMDpattern}), we find that most of the dippers have variable extinction with $A_{V}$ changes in the range of $0.5-1.3$\;mag.

In this section, we have related periodic, burster and dipper categories with spot modulated stellar rotation, variable accretion and extinction changes, respectively. But we should keep in mind that multiple physical processes are taking place in the young star systems, making the above calculations being over simplified. For example, \citet{Rackham2018ApJ...853..122R} pointed out that the spot coverage may be underestimated from lightcurve amplitude alone. Additional high quality observations should be helpful to decouple these physical processes, and to determine the corresponding physical parameters more accurately.

Since YSOs are generally found to be more variable than field stars, time series photometry is powerful in selecting candidate samples dominated by YSOs. Future and existing time-domain surveys, such as the ZTF \citep{Kulkarni2018ATel11266....1K} and the Vera C. Rubin Observatory \citep[Large Synoptic Survey Telescope;][]{Ivezic2019ApJ...873..111I,Bianco2022ApJS..258....1B}, will help search the fainest YSOs that are invisible to astrometry surveys, such as the Gaia \citep{Gaia-Collaboration2016A&A...595A...1G}, as well as further improve our knowledge about the mechanisms related to different variability behavior.

\section{Summary}
\label{sec:summary}

In this work we studied the variability of 288 YSOs in the Perseus molecular cloud using the about 4 years time series data from the ZTF. The main results are summarized as follows.

\begin{enumerate}
    \item We identified 238 sources as variables based on the normalized peak-to-peak variability metric. We found variability fractions of 83\% for the whole working sample, 92\% for the disk population, and 77\% for the diskless population. Disked YSOs are more variable than diskless YSOs.
    \item The variables are classified into 9 morphological categories mainly based the flux asymmetry ($M$) and quasiperiodicity ($Q$) metrics. The dominant variability behavior of these variables are strictly periodic (21.3\% of the variable sample) and quasiperiodic (39.1\% of the variable sample) variables. But for the disk population, the burster, dipper, long timescale and stochastic categories are also common.
    \item We analyze CMD pattern of the variables using quasi-simultaneous multiband photometry from the ZTF.  We found that periodic variables have the steepest CMD pattern, and that bursters have much flatter slopes than dippers.  Periodic and quasiperiodic variables have the lowest variability amplitudes. The periodic variability is consistent with spot modulated stellar rotation, with spot coverage changes of 30\% to 40\%. The burster variability is consistent with accretion induced brightness changes, with accretion luminosity changes in the range of $L_{\rm acc}/L_{\star}=0.1-0.3$. The dipper variability is consistent with variable extinction with $A_{V}$ changes in the range of $0.5-1.5\;\rm mag$.
\end{enumerate}

\begin{acknowledgements}
 We thank the anonymous referee for his/her helpful comments that improved this manuscript. We acknowledge the support of the CAS International Cooperation Program (grant No. 114332KYSB20190009) and the NSFC grant No. 12033004. GJH and XZ are supported by grant 12173003 from the National Natural Science Foundation of China. We thank Prof. Hillenbrand for helpful suggestions. The Guoshoujing Telescope (the Large Sky Area Multi-Object Fiber Spectroscopic Telescope LAMOST) is a National Major Scientific Project built by the Chinese Academy of Sciences. Funding for the project has been provided by the National Development and Reform Commission. LAMOST is operated and managed by the National Astronomical Observatories, Chinese Academy of Sciences. This work is based on observations obtained with the Samuel Oschin Telescope 48-inch and the 60-inch Telescope at the Palomar Observatory as part of the Zwicky Transient Facility project. ZTF is supported by the National Science Foundation under Grants No. AST-1440341 and AST-2034437 and a collaboration including current partners Caltech, IPAC, the Weizmann Institute for Science, the Oskar Klein Center at Stockholm University, the University of Maryland, Deutsches Elektronen-Synchrotron and Humboldt University, the TANGO Consortium of Taiwan, the University of Wisconsin at Milwaukee, Trinity College Dublin, Lawrence Livermore National Laboratories, IN2P3, University of Warwick, Ruhr University Bochum, Northwestern University and former partners the University of Washington, Los Alamos National Laboratories, and Lawrence Berkeley National Laboratories. Operations are conducted by COO, IPAC, and UW. This research made use of \texttt{APLpy}, an open-source plotting package for Python \citep{Robitaille2012aplpy}. This research made use of \texttt{Astropy}, a community-developed core Python package for Astronomy \citep{Astropy-Collaboration2013A&A...558A..33A,Astropy-Collaboration2018AJ....156..123A}. We also acknowledge the various Python packages that were used in the data analysis of this work, including \texttt{Matplotlib} \citep{Hunter2007CSE.....9...90H}, \texttt{NumPy} \citep{Harris2020Natur.585..357H}, \texttt{SciPy} \citep{Virtanen2020NatMe..17..261V}.

\end{acknowledgements}

\appendix%
\section{CMD Pattern Due to Noise}
\label{app:noise}

\begin{figure*}[!t]
    \centering
    \includegraphics[width=\textwidth]{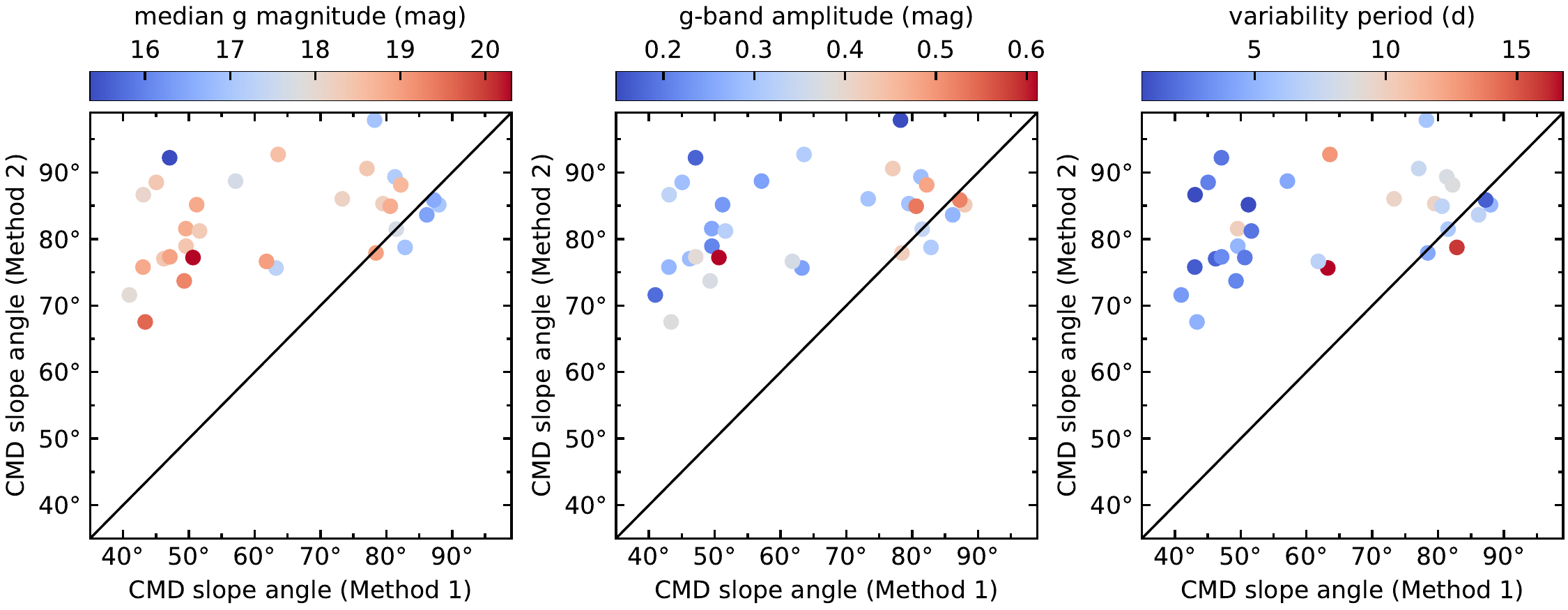}
    \caption{Comparison of CMD slope angles determined using Method~1 and Method~2 for periodic variables in our sample. The symbols are color coded according to the median $g$ magnitude (left), the variability amplitude in $g$-band (middle) and the variability period (right). The straight line in each panel represents the line of equality.}
    \label{fig:angle_comparison}
\end{figure*}

\begin{figure}[!t]
    \centering
    \includegraphics[width=\columnwidth]{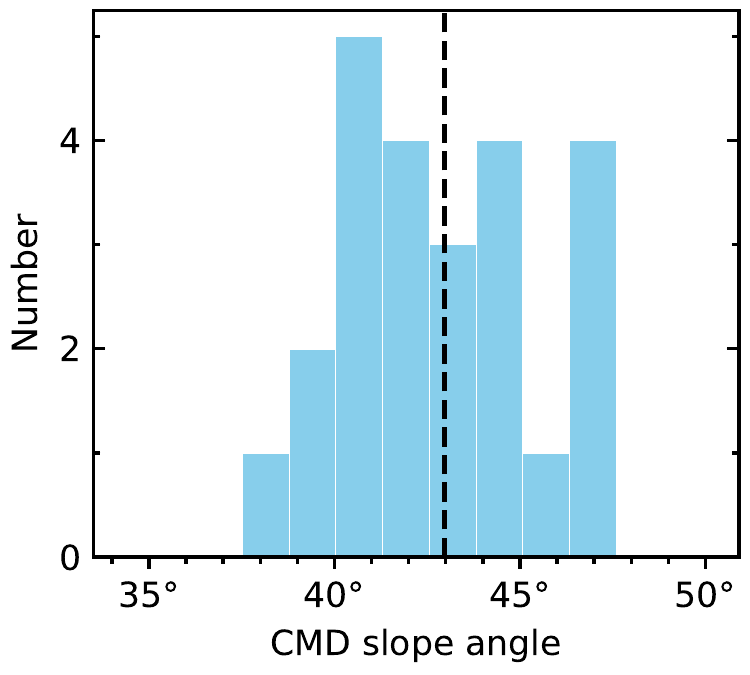}
    \caption{Histogram showing the distribution of CMD slope angles determined using Method~1 for non-variables in our sample. The black vertical line corresponds to the angle of 43$^{\circ}$ for the typical uncertainties of our sample.}
    \label{fig:CMDnoise}
\end{figure}

In Section~\ref{sec:CMDpattern_model}, we note that the CMD slope angles determined using Method 1 for periodic variables deviate significantly from that of spot induced variability, while that from Method~2 is consistent with the model. This discrepancy could be attributed to the noise and the interpolation method. We compare the two angles in Figure~\ref{fig:angle_comparison}, and note trends of decreasing discrepancies between the two angles with increasing brightness, variability amplitudes and variability period. In Figure~\ref{fig:CMDnoise}, we display the CMD slope angles determined using Method 1 for those non-variables, whose CMD pattern should be dominated by the uncertainties in the measurements or introduced during the interpolation for $g$-band photometry. Those non-variables have angles in the range 40$^{\circ}$ to 45$^{\circ}$, in consistent with the peak around 40$^{\circ}$ of the blue histogram in the upper left panel of Figure~\ref{fig:CMDpattern}. In fact, given the typical photometric uncertainties of our sample, 0.06\;mag in $g$-band and 0.02\;mag in $r$-band, a random sampling in the photometry can lead to a CMD slope angle of $\sim$43$^{\circ}$. Some dippers also have CMD slope angles close to $\sim$43$^{\circ}$ in Figure~\ref{fig:CMDpattern}. These sources generally have variability amplitude comparable to the measurement uncertainties and the measurements of their CMD slope angles could be affected by the noise discussed above.

\clearpage%
\onecolumn%
\begin{landscape}
\section{Table List the Properties of Variable}
\small
\setlength{\tabcolsep}{1.7ex}
\renewcommand{\arraystretch}{1.2}
\begin{longtable}{lccccccccccccc}
\caption[]{Properties of Variables in the Perseus Molecular Cloud}\label{tab:variables}\\\toprule
Identifier & R.A.  & Decl. & SPT & Disk$^{a}$ & CTTS$^{b}$ & $\left<r\right>$ & $\nu$ & Amplitude & Timescale & $Q$ & $M$ & Category & Angle\\
           & (deg) & (deg) &     &            &            & (mag)            &       & (mag)     &  (day)    &     &     &          & (deg)\\
\midrule
\endfirsthead
\caption[]{--\,{\it continued}}\\\toprule
Identifier & R.A.  & Decl. & SPT & Disk$^{a}$ & CTTS$^{b}$ & $\left<r\right>$ & $\nu$ & Amplitude & Timescale & $Q$ & $M$ & Category & Angle\\
           & (deg) & (deg) &     &            &            & (mag)            &       & (mag)     &  (day)    &     &     &          & (deg)\\
\midrule
\endhead
\bottomrule
\endfoot
\bottomrule
\multicolumn{14}{l}{\vspace{-2mm}}\\
\multicolumn{14}{l}{$^{a}$ ``Y" for disked objects and ``N" for diskless objects.}\\
\multicolumn{14}{l}{$^{b}$ ``Y" for CTTSs and ``N" for WTTSs, objects without LAMOST spectra are unclassified.}\\
\multicolumn{14}{l}{$^{*}$ The variability types are adjusted to favor our by-eye classification.}
\endlastfoot
Gaia EDR3 124009935063344128  &  51.245429  &  31.010398  &  M3.1  &  N  &         N  &  16.12  &  0.0078  &  0.19  &   50.36  &   0.81  &   0.08  &        QPS  &      68.7\\
Gaia EDR3 123994679339516800  &  51.281136  &  30.828328  &  M2.8  &  N  &         N  &  16.59  &  0.0065  &  0.18  &    3.18  &   0.85  &  -0.04  &        QPS  &      41.0\\
     2MASS J03250943+3046215  &  51.289291  &  30.772661  &  K9.3  &  Y  &         Y  &  18.12  &  0.0299  &  0.96  &    8.31  &   0.92  &  -0.16  &          S  &      62.1\\
Gaia EDR3 123998252752298752  &  51.302425  &  30.989481  &  K9.5  &  Y  &         Y  &  13.29  &  0.0250  &  0.31  &   11.90  &   0.88  &  -0.09  &          S  &      50.9\\
Gaia EDR3 123996874066752640  &  51.379737  &  30.918846  &  M5.2  &  N  &         N  &  17.62  &  0.0090  &  0.33  &   34.59  &   0.65  &  -0.24  &        QPS  &      44.9\\
     2MASS J03253315+3055443  &  51.388165  &  30.928993  &  M0.9  &  Y  &         Y  &  15.74  &  0.0701  &  1.61  &    4.78  &   0.72  &   0.14  &        QPS  &      68.2\\
Gaia EDR3 124018318839496064  &  51.407963  &  31.139139  &  M0.2  &  Y  &         Y  &  15.61  &  0.0576  &  1.10  &    0.86  &   0.93  &   0.45  &        APD  &      78.0\\
Gaia EDR3 123999936379477504  &  51.445275  &  30.955693  &  M4.6  &  Y  &         Y  &  16.83  &  0.0353  &  0.86  &  121.39  &   0.85  &   0.40  &          L  &      71.9\\
     2MASS J03254886+3057258  &  51.453590  &  30.957169  &  K4.5  &  Y  &  $\cdots$  &  14.58  &  0.0516  &  1.03  &  143.26  &   0.93  &   0.53  &          L  &      34.2\\
     2MASS J03255275+3054490  &  51.469818  &  30.913614  &  K5.8  &  Y  &         N  &  15.05  &  0.0184  &  0.28  &    1.17  &   0.88  &   0.68  &        APD  &      79.1\\
Gaia EDR3 124002856957227648  &  51.582532  &  31.110303  &    K7  &  Y  &         N  &  16.09  &  0.0351  &  0.72  &    9.68  &   0.80  &   0.77  &        QPD  &      77.4\\
Gaia EDR3 124030138589484672  &  51.617565  &  31.202155  &    K7  &  Y  &         Y  &  14.47  &  0.0653  &  1.24  &    5.76  &   0.78  &   0.72  &        QPD  &      78.1\\
     2MASS J03274148+3020166  &  51.922852  &  30.337971  &  K8.1  &  Y  &         Y  &  14.61  &  0.0185  &  0.38  &    3.15  &   0.65  &   0.19  &        QPS  &      72.9\\
Gaia EDR3 124034399197073024  &  51.926947  &  31.285487  &  M2.8  &  N  &         N  &  16.23  &  0.0074  &  0.24  &    9.61  &   0.06  &   0.01  &          P  &  $\cdots$\\
Gaia EDR3 124034399197072896  &  51.930860  &  31.288977  &  K4.5  &  N  &         N  &  14.36  &  0.0054  &  0.15  &    0.82  &   0.61  &  -0.41  &          B  &  $\cdots$\\
     2MASS J03282578+3054000  &  52.107437  &  30.900009  &  K2.9  &  N  &         N  &  13.61  &  0.0059  &  0.07  &   45.82  &   0.77  &   0.05  &        QPS  &      38.8\\
     2MASS J03284242+3029530  &  52.176777  &  30.498074  &  M4.5  &  Y  &         Y  &  16.24  &  0.0209  &  0.61  &    5.83  &   0.82  &  -0.65  &          B  &      68.7\\
     2MASS J03284618+3116385  &  52.192458  &  31.277389  &  M4.5  &  Y  &  $\cdots$  &  14.93  &  0.0125  &  0.21  &    3.64  &   0.57  &  -0.39  &  QPS$^{*}$  &      70.0\\
     2MASS J03284782+3116552  &  52.199250  &  31.282000  &  M6.5  &  Y  &  $\cdots$  &  18.36  &  0.0502  &  1.65  &    3.79  &   0.95  &  -0.28  &          B  &      72.4\\
     2MASS J03285105+3116324  &  52.212750  &  31.275694  &    M5  &  Y  &  $\cdots$  &  17.89  &  0.0177  &  0.46  &  115.82  &   0.77  &  -0.46  &          L  &      61.2\\
     2MASS J03285217+3045055  &  52.217407  &  30.751539  &    K4  &  Y  &  $\cdots$  &  12.88  &  0.0327  &  0.44  &    8.16  &   0.56  &   0.22  &  QPD$^{*}$  &      82.7\\
     2MASS J03285290+3116264  &  52.220417  &  31.274056  &    M5  &  Y  &  $\cdots$  &  18.18  &  0.0145  &  0.48  &  211.66  &   0.71  &  -0.33  &          L  &      55.8\\
     2MASS J03290031+3113385  &  52.251292  &  31.227389  &    M3  &  N  &  $\cdots$  &  17.82  &  0.0123  &  0.45  &    3.20  &   0.67  &   0.22  &        QPS  &      44.2\\
     2MASS J03290289+3116010  &  52.262043  &  31.266953  &  K6.2  &  N  &         N  &  15.52  &  0.0059  &  0.11  &    5.20  &   0.74  &  -0.02  &         MP  &      41.8\\
     2MASS J03290406+3117075  &  52.266917  &  31.285417  &  M5.8  &  N  &  $\cdots$  &  18.02  &  0.0082  &  0.33  &   91.31  &   0.71  &  -0.43  &          B  &      44.4\\
     2MASS J03291243+3114565  &  52.301804  &  31.249033  &  K2.4  &  N  &         N  &  14.31  &  0.0051  &  0.08  &    1.03  &   0.84  &   0.13  &        QPS  &      41.8\\
     2MASS J03291465+3133009  &  52.311069  &  31.550257  &    K4  &  N  &         N  &  14.85  &  0.0083  &  0.16  &    5.63  &   0.58  &   0.23  &        QPS  &      81.1\\
     2MASS J03291766+3122451  &  52.323583  &  31.379194  &    M4  &  Y  &  $\cdots$  &  13.59  &  0.0153  &  0.32  &    5.58  &   0.42  &   0.03  &          P  &  $\cdots$\\
     2MASS J03291872+3123254  &  52.328042  &  31.390389  &  M2.5  &  Y  &         N  &  14.59  &  0.0182  &  0.53  &    0.90  &   0.51  &   0.01  &         MP  &  $\cdots$\\
     2MASS J03292187+3115363  &  52.341125  &  31.260083  &    K4  &  Y  &  $\cdots$  &  14.34  &  0.0883  &  2.84  &    2.11  &   0.53  &   0.18  &    P$^{*}$  &      65.9\\
     2MASS J03292314+3120303  &  52.346458  &  31.341778  &  M4.8  &  Y  &         Y  &  16.59  &  0.0153  &  0.42  &    0.72  &   0.60  &   0.34  &        QPD  &  $\cdots$\\
     2MASS J03292349+3123309  &  52.347917  &  31.391944  &  M3.5  &  N  &         N  &  17.34  &  0.0077  &  0.31  &    1.23  &   0.29  &   0.11  &          P  &  $\cdots$\\
     2MASS J03292591+3126401  &  52.358000  &  31.444472  &  G8.5  &  Y  &         Y  &  15.24  &  0.0811  &  2.20  &  119.91  &   0.48  &  -0.99  &          L  &  $\cdots$\\
     2MASS J03292681+3126475  &  52.361708  &  31.446528  &    M2  &  Y  &         N  &  14.10  &  0.0083  &  0.21  &    1.35  &   0.26  &   0.35  &    P$^{*}$  &  $\cdots$\\
     2MASS J03292815+3116285  &  52.367333  &  31.274611  &  M7.5  &  N  &  $\cdots$  &  18.47  &  0.0201  &  0.61  &    2.43  &   0.33  &  -0.32  &    P$^{*}$  &      77.2\\
     2MASS J03292889+3058418  &  52.370375  &  30.978278  &  M4.8  &  Y  &  $\cdots$  &  16.95  &  0.0122  &  0.24  &    1.49  &   0.78  &   0.37  &        QPD  &      58.6\\
     2MASS J03293038+3119034  &  52.376625  &  31.317639  &  M4.2  &  Y  &         Y  &  15.86  &  0.0068  &  0.20  &    8.77  &   0.42  &  -0.04  &          P  &  $\cdots$\\
     2MASS J03293255+3124370  &  52.385667  &  31.410278  &  M4.5  &  Y  &  $\cdots$  &  17.95  &  0.0565  &  1.46  &    2.52  &   0.46  &   0.45  &  QPS$^{*}$  &  $\cdots$\\
     2MASS J03293286+3127126  &  52.386958  &  31.453528  &    M4  &  Y  &         N  &  17.43  &  0.0095  &  0.31  &    2.69  &   0.66  &   0.33  &        QPD  &  $\cdots$\\
    2MASS J03294592+3104406S  &  52.441750  &  31.077500  &    M1  &  Y  &  $\cdots$  &  16.37  &  0.0174  &  0.44  &    0.53  &   0.90  &  -0.39  &          B  &      57.4\\
     2MASS J03295403+3120529  &  52.475125  &  31.348056  &    M4  &  Y  &         Y  &  15.64  &  0.0176  &  0.50  &    3.13  &   0.57  &  -0.03  &         MP  &  $\cdots$\\
     2MASS J03301614+3147595  &  52.567284  &  31.799868  &  K6.3  &  N  &         N  &  14.86  &  0.0208  &  0.52  &    1.33  &   0.36  &   0.27  &    P$^{*}$  &      85.9\\
     2MASS J03302409+3114043  &  52.600375  &  31.234556  &    M5  &  Y  &  $\cdots$  &  17.68  &  0.0163  &  0.50  &    1.77  &   0.86  &  -0.60  &          B  &      58.0\\
     2MASS J03302598+3102179  &  52.608253  &  31.038307  &  K4.5  &  N  &         N  &  13.61  &  0.0080  &  0.17  &    2.23  &   0.28  &   0.06  &          P  &      92.2\\
     2MASS J03303697+3031276  &  52.654041  &  30.524345  &  K4.8  &  Y  &         N  &  17.74  &  0.0589  &  1.24  &  173.10  &   0.93  &   0.78  &          L  &      71.4\\
     2MASS J03304399+3032469  &  52.683304  &  30.546379  &  K5.3  &  Y  &  $\cdots$  &  14.47  &  0.0204  &  0.47  &    1.45  &   0.92  &  -0.23  &          S  &      63.6\\
     2MASS J03311069+3049405  &  52.794563  &  30.827942  &    M4  &  Y  &         Y  &  16.17  &  0.0342  &  1.01  &    7.17  &   0.80  &   0.06  &        QPS  &      70.0\\
     2MASS J03311471+3049554  &  52.811291  &  30.832075  &  K4.2  &  Y  &         Y  &  16.53  &  0.0140  &  0.33  &    6.76  &   0.79  &   0.16  &        QPS  &      88.8\\
     2MASS J03311830+3049395  &  52.826286  &  30.827658  &  K3.8  &  Y  &         Y  &  14.61  &  0.0601  &  1.07  &   39.96  &   0.57  &   0.31  &        QPD  &  $\cdots$\\
     2MASS J03314240+3106249  &  52.926704  &  31.106930  &  K6.6  &  Y  &         Y  &  17.29  &  0.0166  &  0.65  &    0.94  &   0.63  &   0.15  &         MP  &  $\cdots$\\
     2MASS J03323300+3102216  &  53.137524  &  31.039352  &  M4.6  &  Y  &         Y  &  17.05  &  0.0636  &  1.32  &    5.15  &   0.73  &   0.47  &        QPD  &      74.0\\
     2MASS J03323405+3100557  &  53.141911  &  31.015488  &  K5.1  &  Y  &         Y  &  15.53  &  0.0132  &  0.32  &    0.57  &   0.89  &  -0.25  &          B  &      61.5\\
Gaia EDR3 121147871936317184  &  53.372048  &  30.966361  &    K4  &  N  &         N  &  16.87  &  0.0065  &  0.28  &    3.07  &   0.66  &  -0.16  &        QPS  &      46.8\\
     2MASS J03333041+3110504  &  53.376736  &  31.180670  &    K4  &  Y  &         Y  &  13.82  &  0.0301  &  0.64  &    1.03  &   0.93  &  -0.27  &          B  &      64.4\\
     2MASS J03334692+3053500  &  53.445538  &  30.897247  &  M7.1  &  Y  &  $\cdots$  &  18.23  &  0.0152  &  0.53  &    2.07  &   0.90  &  -0.47  &          B  &      49.9\\
     2MASS J03335108+3112278  &  53.462856  &  31.207733  &  M3.9  &  N  &         N  &  16.67  &  0.0082  &  0.28  &    6.73  &   0.59  &  -0.04  &        QPS  &      56.2\\
     2MASS J03340166+3114396  &  53.506939  &  31.244354  &  M0.5  &  Y  &         Y  &  15.37  &  0.0074  &  0.22  &    0.73  &   0.85  &  -0.10  &    B$^{*}$  &      56.0\\
     2MASS J03343079+3113243  &  53.628304  &  31.223419  &    M4  &  Y  &  $\cdots$  &  18.23  &  0.0104  &  0.57  &    1.47  &   0.81  &  -0.10  &        QPS  &      45.8\\
     2MASS J03344987+3115498  &  53.707813  &  31.263859  &  K5.8  &  Y  &         N  &  14.04  &  0.0070  &  0.14  &    1.66  &   0.75  &   0.03  &        QPS  &      46.0\\
Gaia EDR3 120460024334304128  &  54.553027  &  31.074960  &  M3.7  &  N  &         N  &  16.59  &  0.0078  &  0.19  &  123.44  &   0.88  &  -0.12  &          L  &      49.6\\
Gaia EDR3 217317549813446656  &  54.650193  &  31.653468  &  M1.8  &  N  &         N  &  15.35  &  0.0068  &  0.15  &    5.86  &   0.51  &   0.04  &    P$^{*}$  &      97.9\\
Gaia EDR3 120463116710737920  &  54.875402  &  31.110591  &  M4.2  &  N  &         N  &  17.12  &  0.0068  &  0.23  &    0.73  &   0.22  &   0.02  &          P  &      85.2\\
     2MASS J03404040+3137379  &  55.168335  &  31.627199  &  M5.9  &  N  &         N  &  17.97  &  0.0093  &  0.42  &   73.80  &   0.64  &  -0.08  &        QPS  &      48.3\\
Gaia EDR3 217510376665297152  &  55.243254  &  32.506634  &    K5  &  N  &         N  &  13.91  &  0.0085  &  0.14  &    1.46  &   0.64  &   0.06  &        QPS  &      57.3\\
Gaia EDR3 217458493460476928  &  55.310242  &  32.362894  &  M1.1  &  N  &         N  &  16.56  &  0.0162  &  0.48  &    8.64  &   0.11  &   0.51  &    P$^{*}$  &      88.1\\
Gaia EDR3 217347232331662976  &  55.324826  &  32.047493  &  M3.5  &  N  &         N  &  18.06  &  0.0067  &  0.52  &    1.04  &  -2.97  &  -0.24  &          U  &      43.9\\
Gaia EDR3 216527894305651840  &  55.328058  &  31.237570  &  M2.3  &  N  &         N  &  16.02  &  0.0096  &  0.18  &    7.09  &   0.83  &  -0.18  &        QPS  &      64.4\\
     2MASS J03411921+3202037  &  55.330048  &  32.034370  &  K9.2  &  N  &  $\cdots$  &  17.84  &  0.0379  &  0.91  &    1.20  &   0.88  &   0.71  &        APD  &      60.4\\
Gaia EDR3 216585721745264896  &  55.360264  &  31.843630  &    K7  &  N  &         N  &  16.32  &  0.0078  &  0.28  &    2.91  &   0.68  &   0.07  &        QPS  &      44.7\\
     2MASS J03414251+3118567  &  55.427162  &  31.315769  &  K2.5  &  N  &         N  &  13.30  &  0.0120  &  0.13  &    2.94  &   0.62  &   0.27  &        QPD  &      61.1\\
     2MASS J03415745+3148365  &  55.489384  &  31.810162  &  K4.3  &  Y  &         Y  &  15.93  &  0.0112  &  0.32  &    2.24  &   0.25  &  -0.01  &          P  &      81.2\\
Gaia EDR3 217444440327486848  &  55.492348  &  32.238318  &    G6  &  N  &         N  &  13.50  &  0.0055  &  0.15  &    3.87  &   0.64  &  -0.12  &        QPS  &      62.8\\
     2MASS J03422033+3205310  &  55.584732  &  32.091949  &  K7.4  &  Y  &  $\cdots$  &  17.25  &  0.0225  &  0.67  &    7.31  &   0.71  &   0.81  &        QPD  &      69.3\\
Gaia EDR3 216590016712558080  &  55.588347  &  31.953305  &  M2.5  &  N  &         N  &  17.25  &  0.0110  &  0.47  &   71.53  &   0.79  &  -0.24  &    L$^{*}$  &      50.1\\
     2MASS J03422585+3221022  &  55.607746  &  32.350624  &    K6  &  N  &  $\cdots$  &  17.05  &  0.0061  &  0.28  &   17.17  &   0.42  &  -0.16  &          P  &  $\cdots$\\
     2MASS J03422824+3230479  &  55.617699  &  32.513313  &  K7.1  &  N  &         N  &  16.99  &  0.0122  &  0.42  &    4.34  &   0.23  &   0.41  &    P$^{*}$  &      77.9\\
     2MASS J03423219+3143382  &  55.634144  &  31.727291  &  K4.1  &  N  &         N  &  17.27  &  0.0083  &  0.58  &    1.08  &   0.74  &  -0.11  &        QPS  &      46.9\\
     2MASS J03423291+3142205  &  55.637157  &  31.705711  &    K5  &  Y  &         Y  &  17.54  &  0.0376  &  0.79  &    0.98  &   0.86  &   0.79  &        QPD  &      54.7\\
     2MASS J03424360+3159150  &  55.681702  &  31.987524  &    K2  &  N  &         N  &  15.28  &  0.0080  &  0.18  &    0.69  &   0.77  &  -0.03  &        QPS  &      45.7\\
     2MASS J03425467+3143452  &  55.727802  &  31.729233  &  G3.8  &  N  &         N  &  14.38  &  0.0057  &  0.13  &    0.87  &   0.76  &  -0.06  &        QPS  &      56.0\\
     2MASS J03430214+3207276  &  55.758949  &  32.124344  &  M4.2  &  N  &         N  &  17.58  &  0.0078  &  0.38  &   18.28  &   0.85  &  -0.08  &        QPS  &      40.6\\
     2MASS J03430679+3148204  &  55.778292  &  31.805694  &    M3  &  Y  &         N  &  17.26  &  0.0067  &  0.46  &    6.10  &   0.64  &  -0.18  &        QPS  &      43.2\\
     2MASS J03430704+3210182  &  55.779369  &  32.171745  &  K6.2  &  N  &         N  &  16.60  &  0.0074  &  0.24  &    6.16  &   0.45  &   0.05  &          P  &  $\cdots$\\
     2MASS J03431065+3235323  &  55.794384  &  32.592319  &  K5.9  &  N  &         N  &  16.87  &  0.0096  &  0.33  &    5.89  &   0.51  &   0.28  &  QPS$^{*}$  &      68.0\\
     2MASS J03432438+3238316  &  55.851612  &  32.642117  &  K5.9  &  N  &         N  &  16.46  &  0.0063  &  0.28  &    1.32  &   0.37  &   0.07  &          P  &      75.8\\
     2MASS J03432774+3208314  &  55.865625  &  32.142139  &    M4  &  N  &  $\cdots$  &  18.40  &  0.0101  &  0.62  &    1.12  &   4.30  &  -0.25  &          U  &      44.1\\
     2MASS J03432820+3201591  &  55.867583  &  32.033111  &  M1.8  &  Y  &         Y  &  15.56  &  0.0423  &  1.06  &  187.76  &   0.84  &  -0.02  &          L  &      67.6\\
     2MASS J03433205+3206172  &  55.883667  &  32.104833  &  M0.8  &  N  &         N  &  14.92  &  0.0062  &  0.17  &    5.54  &   0.73  &   0.11  &        QPS  &      60.4\\
     2MASS J03433299+3228027  &  55.887466  &  32.467442  &    K8  &  N  &         N  &  17.82  &  0.0088  &  0.54  &   11.82  &   0.38  &  -0.02  &          P  &  $\cdots$\\
Gaia EDR3 217475123573808000  &  55.895627  &  32.526988  &    G5  &  N  &         N  &  13.67  &  0.0074  &  0.13  &    0.85  &   0.60  &   0.23  &        QPS  &      48.5\\
     2MASS J03434461+3208177  &  55.935917  &  32.138306  &  M0.8  &  Y  &         Y  &  15.41  &  0.0135  &  0.29  &   19.52  &   0.68  &   0.07  &        QPS  &      78.1\\
     2MASS J03434788+3217567  &  55.949542  &  32.299139  &  M1.5  &  N  &         N  &  16.14  &  0.0099  &  0.30  &    9.68  &   0.35  &   0.07  &          P  &      86.0\\
     2MASS J03434792+3218461  &  55.949708  &  32.312861  &    M4  &  N  &         N  &  18.19  &  0.0089  &  0.54  &    0.96  &   0.26  &  -0.22  &  QPS$^{*}$  &      46.8\\
     2MASS J03434862+3213507  &  55.952583  &  32.230806  &    M5  &  N  &  $\cdots$  &  18.44  &  0.0090  &  0.54  &    0.64  &   0.39  &  -0.21  &  QPS$^{*}$  &      43.6\\
     2MASS J03434875+3207332  &  55.953167  &  32.125944  &  M1.5  &  N  &         N  &  15.76  &  0.0061  &  0.18  &    0.97  &   0.80  &  -0.02  &        QPS  &      42.7\\
     2MASS J03434881+3215515  &  55.953375  &  32.264361  &  M4.5  &  Y  &         N  &  17.29  &  0.0082  &  0.37  &    2.88  &   0.15  &  -0.06  &          P  &      73.7\\
     2MASS J03434939+3210398  &  55.955792  &  32.177778  &  M3.5  &  N  &  $\cdots$  &  16.64  &  0.0102  &  0.31  &   12.82  &   0.24  &   0.29  &    P$^{*}$  &      92.7\\
Gaia EDR3 216668803593749888  &  55.961863  &  31.905859  &  K3.6  &  N  &         N  &  15.51  &  0.0105  &  0.29  &    1.32  &   0.78  &  -0.20  &        QPS  &      73.8\\
     2MASS J03435141+3231486  &  55.964230  &  32.530193  &  K5.4  &  N  &         N  &  15.81  &  0.0077  &  0.18  &    3.75  &   0.44  &  -0.02  &          P  &      71.6\\
Gaia EDR3 216520163364482560  &  55.972888  &  31.627645  &    K5  &  N  &         N  &  14.03  &  0.0087  &  0.16  &    1.22  &   0.84  &   0.17  &        QPS  &      52.1\\
     2MASS J03435463+3200298  &  55.977625  &  32.008361  &  M4.2  &  N  &         N  &  17.39  &  0.0061  &  0.33  &  179.91  &   0.45  &  -0.19  &          L  &      45.9\\
     2MASS J03435550+3209321  &  55.981292  &  32.159028  &    K0  &  N  &         N  &  13.93  &  0.0069  &  0.17  &    0.83  &   0.72  &   0.22  &        QPS  &      67.2\\
     2MASS J03435602+3202132  &  55.983458  &  32.037028  &    K7  &  Y  &         Y  &  17.65  &  0.0091  &  0.67  &    4.24  &   0.80  &  -0.09  &        QPS  &      48.5\\
     2MASS J03435619+3208362  &  55.984167  &  32.143417  &    M0  &  N  &         N  &  16.55  &  0.0066  &  0.24  &    3.90  &   0.72  &  -0.05  &        QPS  &      44.9\\
     2MASS J03435622+3230178  &  55.984280  &  32.504955  &  K5.2  &  N  &         N  &  16.09  &  0.0086  &  0.20  &    8.19  &   0.56  &   0.13  &        QPS  &      58.3\\
     2MASS J03435856+3217275  &  55.993958  &  32.291028  &  M3.8  &  Y  &         Y  &  16.31  &  0.0218  &  0.66  &   70.43  &   0.89  &  -0.17  &          S  &      63.0\\
     2MASS J03435890+3211270  &  55.995458  &  32.190861  &  M1.8  &  Y  &         Y  &  16.17  &  0.0337  &  1.06  &    6.75  &   0.74  &   0.19  &        QPS  &      61.3\\
     2MASS J03435907+3214213  &  55.996167  &  32.239250  &  M3.5  &  Y  &  $\cdots$  &  17.57  &  0.0671  &  1.05  &   40.14  &   0.94  &   0.12  &          S  &      72.2\\
     2MASS J03435953+3215551  &  55.998125  &  32.265389  &    M1  &  N  &  $\cdots$  &  17.31  &  0.0072  &  0.42  &    7.58  &   0.45  &   0.17  &          P  &  $\cdots$\\
     2MASS J03435970+3214028  &  55.998833  &  32.234222  &  M0.8  &  N  &         N  &  15.80  &  0.0102  &  0.26  &    0.93  &   0.78  &  -0.10  &        QPS  &      76.3\\
     2MASS J03440216+3219399  &  56.009125  &  32.327806  &  M1.5  &  N  &         N  &  15.75  &  0.0115  &  0.34  &    6.10  &   0.28  &   0.38  &    P$^{*}$  &      81.5\\
     2MASS J03440257+3201348  &  56.010792  &  32.026417  &  M4.8  &  N  &         N  &  17.33  &  0.0062  &  0.31  &    1.03  &   0.82  &  -0.03  &        QPS  &      42.2\\
Gaia EDR3 216613999810965760  &  56.014317  &  31.655093  &    K7  &  N  &         N  &  16.55  &  0.0081  &  0.36  &    0.57  &   0.50  &  -0.07  &        QPS  &      55.2\\
     2MASS J03440410+3207170  &  56.017125  &  32.121417  &    M2  &  N  &         N  &  16.76  &  0.0071  &  0.25  &   10.06  &   0.35  &   0.23  &          P  &      81.6\\
Gaia EDR3 216723568719908992  &  56.019145  &  32.468395  &  M3.3  &  N  &         N  &  18.12  &  0.0162  &  0.53  &    7.00  &   0.52  &  -0.03  &        QPS  &      44.0\\
     2MASS J03440499+3209537  &  56.020833  &  32.164944  &  K3.5  &  N  &         N  &  14.36  &  0.0065  &  0.13  &   22.15  &   0.79  &  -0.02  &        QPS  &      53.5\\
     2MASS J03440646+3143250  &  56.026955  &  31.723621  &  G0.4  &  N  &         N  &  13.99  &  0.0066  &  0.13  &    2.54  &   0.87  &   0.04  &        QPS  &      37.5\\
     2MASS J03440678+3207540  &  56.028292  &  32.131694  &  M4.2  &  Y  &  $\cdots$  &  17.65  &  0.0072  &  0.43  &  179.91  &   0.74  &  -0.17  &          L  &      45.0\\
     2MASS J03440750+3204088  &  56.031292  &  32.069139  &  M4.8  &  Y  &  $\cdots$  &  18.14  &  0.0101  &  0.57  &   76.79  &   0.88  &  -0.43  &          B  &      51.8\\
     2MASS J03440885+3216105  &  56.036917  &  32.269639  &    K0  &  N  &         N  &  14.49  &  0.0084  &  0.14  &   79.72  &   0.89  &  -0.07  &          S  &      50.2\\
     2MASS J03441122+3208161  &  56.046750  &  32.137861  &  M5.2  &  N  &         N  &  17.81  &  0.0106  &  0.40  &    0.68  &   0.79  &  -0.16  &         MP  &      38.8\\
     2MASS J03441125+3206121  &  56.046917  &  32.103361  &    M0  &  N  &         N  &  16.09  &  0.0080  &  0.22  &   10.47  &   0.48  &   0.04  &        QPS  &      50.1\\
     2MASS J03441143+3219401  &  56.047667  &  32.327806  &    M3  &  N  &         N  &  16.30  &  0.0069  &  0.21  &    3.31  &   0.68  &  -0.28  &          B  &      49.2\\
     2MASS J03441568+3231282  &  56.065353  &  32.524506  &    K5  &  N  &         N  &  14.57  &  0.0059  &  0.11  &    1.08  &   0.86  &   0.03  &        QPS  &      53.7\\
     2MASS J03441586+3218396  &  56.066125  &  32.310972  &    M4  &  N  &  $\cdots$  &  17.98  &  0.0112  &  0.57  &   72.03  &   0.89  &  -0.26  &          B  &      47.7\\
     2MASS J03441642+3209552  &  56.068458  &  32.165333  &    K0  &  N  &         N  &  13.57  &  0.0061  &  0.16  &    1.50  &   0.58  &   0.01  &        QPS  &      54.9\\
     2MASS J03441791+3212203  &  56.074625  &  32.205667  &  M2.5  &  N  &         N  &  15.83  &  0.0123  &  0.24  &    4.39  &   0.40  &   0.12  &          P  &      88.7\\
IC 348 IRS J03441827+3207325  &  56.076125  &  32.125694  &  M4.8  &  Y  &  $\cdots$  &  18.69  &  0.0213  &  3.06  &    1.48  &   0.87  &  -0.72  &          B  &  $\cdots$\\
IC 348 IRS J03441925+3207347  &  56.080208  &  32.126306  &  M3.8  &  Y  &         N  &  16.98  &  0.0431  &  0.90  &    0.69  &   0.95  &   0.90  &        APD  &  $\cdots$\\
     2MASS J03442001+3206455  &  56.083417  &  32.112667  &  M3.5  &  N  &  $\cdots$  &  18.18  &  0.0151  &  0.78  &    8.69  &   0.15  &   0.10  &          P  &  $\cdots$\\
     2MASS J03442017+3208565  &  56.084083  &  32.149056  &    M2  &  Y  &  $\cdots$  &  17.04  &  0.0123  &  0.34  &  223.97  &   0.92  &  -0.30  &          L  &      45.3\\
     2MASS J03442023+3230228  &  56.084332  &  32.506344  &  K9.6  &  N  &  $\cdots$  &  17.53  &  0.0550  &  2.73  &    1.89  &   0.96  &   0.15  &          S  &  $\cdots$\\
     2MASS J03442125+3205024  &  56.088583  &  32.084000  &  M2.5  &  N  &  $\cdots$  &  17.10  &  0.0105  &  0.38  &    6.87  &   0.67  &   0.08  &        QPS  &      49.2\\
     2MASS J03442155+3210174  &  56.089833  &  32.171500  &  M1.5  &  N  &         N  &  16.67  &  0.0163  &  0.53  &    7.09  &   0.17  &   0.28  &    P$^{*}$  &      84.9\\
     2MASS J03442156+3215098  &  56.089833  &  32.252722  &  M4.8  &  Y  &         Y  &  17.43  &  0.0153  &  0.42  &    8.14  &   0.73  &   0.10  &        QPS  &      52.5\\
     2MASS J03442161+3210376  &  56.090042  &  32.177111  &    K7  &  Y  &  $\cdots$  &  16.90  &  0.0551  &  1.48  &    7.40  &   0.88  &   0.14  &  QPS$^{*}$  &      69.0\\
     2MASS J03442166+3206248  &  56.090250  &  32.106889  &  M2.8  &  N  &  $\cdots$  &  16.56  &  0.0122  &  0.28  &    8.37  &   0.80  &   0.03  &        QPS  &      63.2\\
     2MASS J03442176+3212312  &  56.090667  &  32.208722  &  M3.5  &  N  &  $\cdots$  &  17.22  &  0.0057  &  0.26  &    3.26  &   0.69  &  -0.05  &        QPS  &      42.7\\
     2MASS J03442186+3217273  &  56.091083  &  32.290917  &  M4.8  &  Y  &  $\cdots$  &  17.96  &  0.0515  &  0.44  &  151.36  &   0.87  &  -0.80  &          L  &  $\cdots$\\
     2MASS J03442191+3212115  &  56.091292  &  32.203222  &    M4  &  N  &  $\cdots$  &  16.76  &  0.0073  &  0.23  &    0.93  &   0.83  &  -0.06  &         MP  &      46.3\\
     2MASS J03442228+3205427  &  56.092875  &  32.095194  &    K8  &  Y  &  $\cdots$  &  16.59  &  0.0338  &  0.59  &  114.91  &   0.80  &  -0.41  &          L  &      95.3\\
     2MASS J03442232+3212007  &  56.093000  &  32.200222  &    M1  &  Y  &         Y  &  16.43  &  0.0245  &  0.90  &  192.20  &   0.74  &  -0.69  &          L  &      69.7\\
     2MASS J03442257+3201536  &  56.094042  &  32.031583  &  M2.5  &  Y  &         N  &  16.33  &  0.0109  &  0.28  &   44.67  &   0.49  &  -0.12  &        QPS  &      60.2\\
     2MASS J03442297+3211572  &  56.095750  &  32.199250  &  M2.2  &  N  &  $\cdots$  &  16.38  &  0.0065  &  0.21  &    5.06  &   0.18  &  -0.05  &          P  &      78.9\\
     2MASS J03442356+3209338  &  56.098208  &  32.159444  &    M5  &  Y  &  $\cdots$  &  18.33  &  0.0099  &  0.61  &    2.13  &   0.86  &  -0.31  &          B  &      45.9\\
     2MASS J03442366+3206465  &  56.098625  &  32.112944  &  M2.5  &  N  &  $\cdots$  &  16.21  &  0.0113  &  0.29  &    9.71  &   0.29  &  -0.06  &          P  &      85.3\\
     2MASS J03442398+3211000  &  56.099958  &  32.183333  &    G0  &  N  &  $\cdots$  &  12.89  &  0.0079  &  0.12  &    0.62  &   0.86  &   0.00  &        QPS  &      45.3\\
     2MASS J03442457+3203571  &  56.102375  &  32.065861  &    M1  &  N  &  $\cdots$  &  17.38  &  0.0124  &  0.42  &    4.94  &   0.55  &   0.12  &    P$^{*}$  &  $\cdots$\\
     2MASS J03442457+3210030  &  56.102417  &  32.167472  &  M3.8  &  N  &  $\cdots$  &  18.57  &  0.0338  &  0.92  &    1.16  &   0.96  &  -0.40  &          B  &      47.7\\
     2MASS J03442557+3212299  &  56.106542  &  32.208333  &  M0.5  &  N  &         N  &  15.23  &  0.0126  &  0.29  &    8.38  &   0.36  &  -0.47  &    P$^{*}$  &      89.3\\
     2MASS J03442602+3204304  &  56.108458  &  32.075111  &    G8  &  Y  &  $\cdots$  &  14.17  &  0.0272  &  0.45  &   71.67  &   0.84  &   0.69  &        QPD  &      70.1\\
     2MASS J03442663+3203583  &  56.110958  &  32.066194  &  M4.8  &  N  &  $\cdots$  &  16.71  &  0.0098  &  0.36  &    3.09  &   0.59  &   0.05  &  QPD$^{*}$  &      53.0\\
     2MASS J03442702+3204436  &  56.112583  &  32.078778  &    M1  &  N  &         N  &  15.19  &  0.0061  &  0.16  &    8.99  &   0.55  &  -0.03  &        QPS  &      57.0\\
     2MASS J03442724+3214209  &  56.113542  &  32.239167  &  M3.5  &  Y  &  $\cdots$  &  18.31  &  0.0844  &  2.02  &    0.51  &   0.92  &   0.17  &          S  &      92.3\\
     2MASS J03442766+3233495  &  56.115276  &  32.563755  &  K6.4  &  N  &         N  &  16.26  &  0.0097  &  0.28  &    1.59  &   0.17  &   0.03  &          P  &      77.0\\
     2MASS J03442787+3207316  &  56.116167  &  32.125444  &    M2  &  N  &  $\cdots$  &  16.11  &  0.0062  &  0.18  &    6.55  &   0.52  &   0.06  &    P$^{*}$  &  $\cdots$\\
     2MASS J03442789+3227189  &  56.116250  &  32.455194  &    M4  &  Y  &         Y  &  17.68  &  0.0340  &  1.18  &   30.64  &   0.89  &   0.00  &          S  &      61.7\\
     2MASS J03442812+3216002  &  56.117167  &  32.266750  &  M3.2  &  N  &  $\cdots$  &  16.32  &  0.0096  &  0.29  &    2.67  &   0.13  &   0.07  &          P  &      88.5\\
     2MASS J03442847+3207224  &  56.118625  &  32.122889  &  K6.5  &  N  &  $\cdots$  &  14.65  &  0.0121  &  0.27  &    6.99  &   0.21  &  -0.05  &          P  &      83.6\\
     2MASS J03442851+3159539  &  56.118792  &  31.998361  &  M3.5  &  Y  &         N  &  16.35  &  0.0070  &  0.26  &   25.25  &   0.70  &   0.19  &        QPS  &      60.5\\
     2MASS J03442912+3207573  &  56.121333  &  32.132611  &  M4.5  &  N  &  $\cdots$  &  18.16  &  0.0072  &  0.45  &    0.80  &   0.69  &   0.00  &        QPS  &      44.1\\
     2MASS J03442972+3210398  &  56.123833  &  32.177722  &    K8  &  Y  &  $\cdots$  &  15.91  &  0.0173  &  0.51  &  127.75  &   0.83  &  -0.01  &          L  &      65.8\\
     2MASS J03442997+3219227  &  56.124917  &  32.322972  &    M4  &  Y  &         N  &  17.50  &  0.0101  &  0.48  &    8.68  &   0.14  &  -0.23  &          P  &  $\cdots$\\
     2MASS J03443054+3206297  &  56.127250  &  32.108250  &    M2  &  N  &  $\cdots$  &  16.81  &  0.0450  &  1.56  &    3.52  &   0.91  &  -0.68  &          B  &      43.2\\
     2MASS J03443153+3208449  &  56.131375  &  32.145833  &    K2  &  N  &  $\cdots$  &  13.25  &  0.0071  &  0.14  &    2.24  &   0.60  &  -0.08  &        QPS  &      40.4\\
     2MASS J03443259+3208424  &  56.135792  &  32.145139  &  M2.5  &  N  &  $\cdots$  &  15.17  &  0.0065  &  0.15  &  110.73  &   0.79  &   0.08  &          L  &      38.7\\
     2MASS J03443274+3208374  &  56.136417  &  32.143750  &    G6  &  N  &  $\cdots$  &  12.86  &  0.0081  &  0.14  &    2.60  &   0.83  &   0.01  &        QPS  &      42.9\\
     2MASS J03443276+3209157  &  56.136542  &  32.154389  &  M3.2  &  Y  &  $\cdots$  &  16.87  &  0.0082  &  0.37  &    5.28  &   0.43  &  -0.09  &          P  &  $\cdots$\\
     2MASS J03443321+3215290  &  56.138417  &  32.258083  &  M2.2  &  N  &  $\cdots$  &  16.21  &  0.0062  &  0.18  &    1.04  &   0.84  &  -0.13  &        QPS  &      44.8\\
     2MASS J03443330+3209396  &  56.138792  &  32.161000  &    M2  &  Y  &  $\cdots$  &  16.36  &  0.0402  &  1.50  &    0.90  &   0.93  &   0.32  &        APD  &      40.3\\
     2MASS J03443398+3208541  &  56.141583  &  32.148361  &    M0  &  N  &         N  &  15.02  &  0.0143  &  0.31  &   15.88  &   0.41  &   0.13  &          P  &      78.7\\
     2MASS J03443426+3210497  &  56.142792  &  32.180472  &    M2  &  N  &  $\cdots$  &  16.73  &  0.0065  &  0.29  &    1.76  &   0.78  &  -0.18  &        QPS  &      45.1\\
     2MASS J03443481+3156552  &  56.145042  &  31.948667  &  M2.5  &  Y  &         Y  &  17.14  &  0.0120  &  0.43  &    3.48  &   0.59  &   0.10  &        QPS  &  $\cdots$\\
     2MASS J03443487+3206337  &  56.145333  &  32.109333  &  K5.5  &  N  &  $\cdots$  &  14.75  &  0.0086  &  0.22  &    5.47  &   0.23  &  -0.54  &    P$^{*}$  &  $\cdots$\\
     2MASS J03443498+3215311  &  56.145792  &  32.258639  &  M3.5  &  Y  &  $\cdots$  &  18.12  &  0.0089  &  0.38  &    1.03  &   0.46  &  -0.52  &          B  &  $\cdots$\\
     2MASS J03443503+3207370  &  56.146000  &  32.126917  &  K6.5  &  Y  &  $\cdots$  &  14.35  &  0.0049  &  0.18  &    4.54  &   0.25  &   0.29  &    P$^{*}$  &  $\cdots$\\
     2MASS J03443537+3207362  &  56.147375  &  32.126694  &    M0  &  Y  &  $\cdots$  &  16.40  &  0.0288  &  0.54  &    0.88  &   0.34  &   0.40  &  QPD$^{*}$  &  $\cdots$\\
     2MASS J03443568+3203035  &  56.148708  &  32.051000  &  M3.2  &  Y  &  $\cdots$  &  18.34  &  0.0413  &  1.48  &    1.88  &   0.44  &   0.67  &        EYE  &  $\cdots$\\
     2MASS J03443741+3209009  &  56.155875  &  32.150250  &    M1  &  Y  &         N  &  17.03  &  0.0503  &  0.88  &    0.89  &   0.72  &   0.81  &        QPD  &  $\cdots$\\
     2MASS J03443788+3208041  &  56.157875  &  32.134500  &    K7  &  Y  &  $\cdots$  &  16.17  &  0.0203  &  0.77  &    0.89  &   0.78  &  -0.19  &        QPS  &      64.3\\
     2MASS J03443798+3203296  &  56.158292  &  32.058278  &    K6  &  Y  &         Y  &  15.36  &  0.0676  &  1.54  &    1.13  &   0.64  &   0.13  &        QPS  &      81.4\\
     2MASS J03443800+3211370  &  56.158375  &  32.193639  &    M4  &  Y  &  $\cdots$  &  18.02  &  0.0100  &  0.67  &  250.00  &   0.59  &   0.23  &          L  &  $\cdots$\\
     2MASS J03443845+3207356  &  56.160292  &  32.126583  &    K6  &  Y  &  $\cdots$  &  14.88  &  0.0163  &  0.43  &    5.22  &   0.09  &   0.23  &          P  &      85.1\\
     2MASS J03443854+3208006  &  56.160625  &  32.133528  &  M1.2  &  Y  &         Y  &  16.04  &  0.0148  &  0.42  &    7.52  &   0.19  &   0.21  &          P  &      90.6\\
     2MASS J03443878+3219056  &  56.161625  &  32.318222  &  M3.5  &  N  &  $\cdots$  &  17.55  &  0.0089  &  0.38  &    4.84  &   0.27  &  -0.09  &          P  &      67.5\\
     2MASS J03443916+3209182  &  56.163250  &  32.155111  &    G8  &  N  &         N  &  13.78  &  0.0069  &  0.17  &    0.62  &   0.71  &  -0.19  &        QPS  &      43.4\\
     2MASS J03443919+3209448  &  56.163375  &  32.162417  &    M2  &  Y  &  $\cdots$  &  17.04  &  0.0087  &  0.45  &    0.80  &   0.40  &  -0.06  &          P  &  $\cdots$\\
     2MASS J03443924+3207355  &  56.163542  &  32.126528  &    K3  &  N  &  $\cdots$  &  15.13  &  0.0269  &  0.85  &  118.63  &   0.64  &   0.72  &          L  &      88.2\\
     2MASS J03443979+3218041  &  56.165875  &  32.301167  &  M3.8  &  Y  &         Y  &  16.99  &  0.0145  &  0.44  &    7.90  &   0.48  &  -0.26  &          B  &  $\cdots$\\
     2MASS J03443985+3215580  &  56.166083  &  32.266139  &  M3.5  &  N  &  $\cdots$  &  18.28  &  0.0089  &  0.50  &    1.45  &   0.14  &  -0.04  &          P  &  $\cdots$\\
     2MASS J03444011+3211341  &  56.167208  &  32.192861  &    K2  &  N  &         N  &  15.05  &  0.0070  &  0.16  &    2.13  &   0.50  &  -0.05  &        QPS  &  $\cdots$\\
     2MASS J03444061+3223110  &  56.169209  &  32.386414  &  K5.2  &  N  &         N  &  16.85  &  0.0069  &  0.25  &    0.80  &   0.11  &   0.08  &          P  &  $\cdots$\\
     2MASS J03444207+3209009  &  56.175083  &  32.150028  &  M4.2  &  Y  &  $\cdots$  &  17.29  &  0.0159  &  0.57  &    0.60  &   0.81  &  -0.53  &          B  &      49.7\\
IC 348 IRS J03444215+3209022  &  56.175625  &  32.150611  &  M2.5  &  Y  &  $\cdots$  &  17.51  &  0.0424  &  1.15  &    0.97  &   0.78  &   0.19  &         MP  &      66.6\\
     2MASS J03444261+3206194  &  56.177625  &  32.105417  &    M1  &  N  &  $\cdots$  &  15.30  &  0.0077  &  0.19  &   11.64  &   0.47  &  -0.05  &        QPS  &  $\cdots$\\
     2MASS J03444351+3207427  &  56.181333  &  32.128611  &    M1  &  N  &  $\cdots$  &  17.06  &  0.0096  &  0.49  &    1.14  &   0.65  &   0.17  &        QPS  &      46.0\\
     2MASS J03444376+3210304  &  56.182417  &  32.175167  &  M1.2  &  Y  &  $\cdots$  &  16.35  &  0.0488  &  1.61  &    1.15  &   0.85  &   0.61  &        QPD  &      70.8\\
     2MASS J03444458+3208125  &  56.185792  &  32.136861  &    M2  &  Y  &  $\cdots$  &  17.44  &  0.0290  &  0.82  &    1.39  &   0.80  &   0.16  &        QPS  &      47.1\\
     2MASS J03444472+3204024  &  56.186333  &  32.067417  &  M0.5  &  Y  &         Y  &  15.17  &  0.0198  &  0.65  &    0.55  &   0.86  &  -0.31  &          B  &      64.9\\
     2MASS J03444495+3213364  &  56.187417  &  32.226833  &  M4.8  &  N  &  $\cdots$  &  17.89  &  0.0098  &  0.50  &    1.06  &   0.82  &  -0.20  &         MP  &      47.2\\
     2MASS J03444881+3213218  &  56.203458  &  32.222806  &  M2.8  &  N  &         N  &  17.01  &  0.0104  &  0.37  &    6.90  &   0.31  &   0.26  &    P$^{*}$  &      76.6\\
     2MASS J03445096+3216093  &  56.212375  &  32.269333  &  M3.2  &  N  &         N  &  16.67  &  0.0111  &  0.27  &   12.15  &   0.60  &   0.09  &        QPS  &      49.4\\
     2MASS J03445274+3200565  &  56.219792  &  32.015778  &  M4.8  &  N  &  $\cdots$  &  17.22  &  0.0075  &  0.30  &   10.22  &   0.83  &  -0.36  &          B  &      44.8\\
     2MASS J03445561+3209198  &  56.231792  &  32.155611  &    K4  &  N  &         N  &  16.34  &  0.0086  &  0.39  &    3.14  &   0.18  &  -0.01  &          P  &      77.3\\
     2MASS J03445611+3205564  &  56.233833  &  32.099083  &  M2.8  &  N  &  $\cdots$  &  16.80  &  0.0071  &  0.22  &    3.31  &   0.66  &  -0.13  &        QPS  &      38.7\\
     2MASS J03445614+3209152  &  56.233958  &  32.154306  &    K0  &  Y  &         Y  &  14.77  &  0.0287  &  0.63  &  230.32  &   0.77  &   1.00  &          L  &      80.5\\
     2MASS J03445785+3204016  &  56.241083  &  32.067167  &  M5.2  &  Y  &  $\cdots$  &  18.33  &  0.0096  &  0.49  &    2.30  &   0.51  &  -0.07  &        QPS  &      46.1\\
     2MASS J03450108+3203200  &  56.254542  &  32.055611  &  M4.2  &  N  &  $\cdots$  &  18.10  &  0.0080  &  0.47  &    1.03  &   0.72  &  -0.12  &        QPS  &      46.0\\
     2MASS J03450148+3212288  &  56.256167  &  32.208083  &    M4  &  N  &  $\cdots$  &  17.66  &  0.0068  &  0.41  &    2.06  &   0.61  &  -0.08  &        QPS  &      45.4\\
     2MASS J03450151+3210512  &  56.256333  &  32.180972  &    K0  &  N  &         N  &  14.24  &  0.0073  &  0.14  &    1.88  &   0.69  &  -0.08  &        QPS  &      44.3\\
     2MASS J03450174+3214276  &  56.257250  &  32.241083  &    K4  &  N  &         N  &  15.11  &  0.0090  &  0.24  &   16.76  &   0.46  &   0.22  &    P$^{*}$  &      75.6\\
     2MASS J03450285+3207006  &  56.261875  &  32.116917  &  M4.8  &  N  &         N  &  17.93  &  0.0088  &  0.51  &    0.98  &   0.80  &  -0.17  &        QPS  &      46.3\\
     2MASS J03450521+3209544  &  56.271750  &  32.165139  &    M3  &  N  &  $\cdots$  &  17.72  &  0.0077  &  0.45  &    4.95  &   0.74  &   0.23  &        QPS  &      44.4\\
     2MASS J03450577+3203080  &  56.274042  &  32.052278  &    M0  &  N  &         N  &  14.60  &  0.0076  &  0.14  &    0.96  &   0.85  &  -0.02  &        QPS  &  $\cdots$\\
     2MASS J03450773+3200272  &  56.282230  &  32.007576  &  K2.1  &  N  &         N  &  12.61  &  0.0168  &  0.22  &    1.57  &   0.90  &  -0.19  &  QPS$^{*}$  &      42.9\\
     2MASS J03450796+3204018  &  56.283167  &  32.067250  &    G4  &  N  &         N  &  14.44  &  0.0062  &  0.13  &    1.04  &   0.76  &   0.17  &        QPS  &      34.3\\
     2MASS J03451598+3230519  &  56.316608  &  32.514442  &  K2.9  &  N  &         N  &  15.32  &  0.0069  &  0.16  &    0.57  &   0.80  &   0.15  &        QPS  &      36.4\\
     2MASS J03451634+3206199  &  56.318125  &  32.105528  &    K5  &  Y  &         Y  &  16.22  &  0.0620  &  1.22  &    1.46  &   0.93  &   0.73  &        APD  &      73.9\\
     2MASS J03451782+3212058  &  56.324292  &  32.201639  &  M3.8  &  Y  &  $\cdots$  &  18.32  &  0.0148  &  0.73  &  198.99  &   0.48  &  -0.53  &          L  &      51.9\\
     2MASS J03451799+3219330  &  56.325000  &  32.325833  &  M3.5  &  N  &         N  &  17.99  &  0.0136  &  0.62  &    7.67  &   0.74  &  -0.29  &  QPS$^{*}$  &      44.6\\
     2MASS J03452046+3206344  &  56.335250  &  32.109556  &    M1  &  Y  &         N  &  15.19  &  0.0206  &  0.41  &    3.16  &   0.86  &   0.54  &        QPD  &      69.0\\
Gaia EDR3 216729139294155520  &  56.340596  &  32.535158  &  M3.6  &  N  &         N  &  17.57  &  0.0082  &  0.43  &    5.08  &   0.69  &  -0.08  &        QPS  &      45.4\\
     2MASS J03452214+3202040  &  56.342292  &  32.034444  &    M4  &  N  &         N  &  17.38  &  0.0057  &  0.31  &    0.97  &   0.79  &  -0.13  &         MP  &      43.6\\
Gaia EDR3 216714158448241792  &  56.347417  &  32.410264  &  K5.6  &  N  &         N  &  15.77  &  0.0073  &  0.18  &    8.51  &   0.76  &   0.36  &        QPD  &      43.6\\
     2MASS J03452514+3209301  &  56.354792  &  32.158389  &  M3.8  &  Y  &         Y  &  16.52  &  0.0184  &  0.79  &  119.02  &   0.83  &  -0.91  &          L  &      60.2\\
     2MASS J03453061+3201557  &  56.377542  &  32.032111  &    K6  &  N  &         N  &  14.49  &  0.0070  &  0.14  &    2.24  &   0.68  &  -0.04  &        QPS  &      49.4\\
     2MASS J03453230+3203150  &  56.384583  &  32.054139  &    M3  &  N  &  $\cdots$  &  16.33  &  0.0108  &  0.33  &    0.71  &   0.08  &  -0.50  &    P$^{*}$  &      86.7\\
     2MASS J03453545+3203259  &  56.397708  &  32.057167  &    M4  &  N  &  $\cdots$  &  17.10  &  0.0080  &  0.26  &    1.57  &   0.49  &  -0.06  &    P$^{*}$  &  $\cdots$\\
     2MASS J03453551+3156257  &  56.397968  &  31.940475  &    K8  &  N  &         N  &  16.94  &  0.0171  &  0.31  &    9.30  &   0.88  &  -0.20  &  QPS$^{*}$  &      25.7\\
     2MASS J03454675+3228487  &  56.444813  &  32.480206  &  K7.5  &  N  &         N  &  17.42  &  0.0109  &  0.55  &   10.42  &   0.53  &  -0.17  &    P$^{*}$  &  $\cdots$\\
Gaia EDR3 216418321101566464  &  56.521716  &  31.647741  &  M0.3  &  N  &         N  &  16.53  &  0.0065  &  0.28  &    7.46  &   0.58  &  -0.10  &        QPS  &      49.6\\
Gaia EDR3 216643480466527744  &  56.693603  &  32.030345  &    K7  &  N  &         N  &  15.04  &  0.0072  &  0.14  &    4.44  &   0.83  &   0.01  &        QPS  &      47.6\\
     2MASS J03465739+3249173  &  56.739124  &  32.821499  &  M5.3  &  Y  &  $\cdots$  &  18.54  &  0.0089  &  0.71  &   30.06  &   0.70  &  -0.12  &        QPS  &      42.5\\
     2MASS J03474711+3304034  &  56.946308  &  33.067616  &  K4.9  &  Y  &         Y  &  15.41  &  0.0242  &  0.32  &  118.25  &   0.82  &   0.97  &          L  &      79.2\\
     2MASS J03480048+3207463  &  57.002022  &  32.129539  &  M2.9  &  Y  &  $\cdots$  &  17.89  &  0.0134  &  0.45  &    1.58  &   0.82  &   0.55  &        QPD  &      45.6\\
Gaia EDR3 216829401012198912  &  57.012845  &  32.086724  &    K7  &  N  &         N  &  14.12  &  0.0072  &  0.17  &  183.98  &   0.85  &  -0.02  &          L  &      76.6\\
Gaia EDR3 217072186920843264  &  57.213927  &  32.714366  &  M2.2  &  N  &         N  &  15.30  &  0.0061  &  0.12  &    1.29  &   0.74  &   0.18  &        QPS  &  $\cdots$\\
Gaia EDR3 217068063754243584  &  57.453438  &  32.833972  &  M2.8  &  N  &         N  &  16.73  &  0.0081  &  0.24  &    4.69  &   0.67  &   0.21  &        QPS  &      48.8
\end{longtable}
\end{landscape}
\twocolumn%

\label{lastpage}

\bibliographystyle{raa}
\bibliography{ms2023-0099}

\end{document}